\documentclass[a4paper,fleqn]{cas-dc}

\usepackage[numbers]{natbib}

\def\tsc#1{\csdef{#1}{\textsc{\lowercase{#1}}\xspace}}
\tsc{WGM}
\tsc{QE}
\tsc{EP}
\tsc{PMS}
\tsc{BEC}
\tsc{DE}

\begin{document}
\let\WriteBookmarks\relax
\def\floatpagepagefraction{1}
\def\textpagefraction{.001}
\shortauthors{S. Sharma et~al.}

\title [mode = title]{Effect of geometric parameters on the noise generated by rod-airfoil configuration}                      



\author[1]{Sparsh Sharma}[type=editor,
                        auid=000,bioid=1,
                        prefix=,
                        role=,
                        orcid=0000-0002-6978-8663]
\cormark[1]
\ead{sparsh.sharma@b-tu.de}

\credit{Conceptualization, Methodology, Data curation, Writing - Original draft preparation}
\address[1]{Technische Akustik, Brandenburgische Technische Universität Cottbus-Senftenberg, Cottbus, Germany}

\author[1]{Thomas F. Geyer}

\credit{Conceptualization, Methodology, Data curation, Writing - Original draft preparation}

\author[1]{Jens Giesler}

\credit{Conceptualization, Methodology, Data curation, Writing - Original draft preparation}










\cortext[cor1]{Corresponding author}


\begin{abstract}
This paper investigates the effect of the geometric parameters -- rod diameter and the distance between the rod and the airfoil -- on the noise generated by rod-airfoil configuration using experimental and numerical techniques. The numerical simulations are carried out using the low-dissipation up-wind scheme and the Delayed-Detached-Eddy Simulation (DDES) approach with Shear-Layer Adapted (SLA) sub-grid length scale (SGS) for a faster transition between RANS and LES. A dual-time stepping strategy, leading to second-order accuracy in space and time, is followed. The Ffowcs-Williams and Hawkings (FWH) technique is used to post-process the pressure fluctuations to predict the far~field acoustics. A corresponding detailed experimental analysis, carried out using phased microphone array techniques in the aeroacoustic wind tunnel at the Brandenburg Technical University at Cottbus, is used for the validation of the numerical method. The key objective of the analysis is to examine the influence of the parameters of the configuration on the noise generation. The results show reasonable agreement with the experimental data in terms of far-field acoustics.
\end{abstract}



\begin{keywords}
rod-airfoil \sep broadband noise \sep DDES \sep FW-H \sep microphone array
\end{keywords}

\maketitle

\section{Introduction}
An airfoil subjected to a real flow emits noise that mainly consists of two components: the first is the leading edge (LE) noise, which results due to the interaction of the leading edge of the airfoil and the turbulent inflow, and the second one is the trailing edge noise, which results due the interaction of the airfoil boundary layer with the trailing edge. The former becomes an important broadband noise generating mechanism in terms of unsteady loading at reasonable angles of attack and highly perturbed flow \citep{Blake1986}. As such it is observed in many practical applications like turbomachinery due to the interaction between the rotor wake and the leading edges of the downstream located stator blades \citep{Schulten1997,Cooper2005,Cooper2006}. To investigate the generation of such a phenomenon, a rod-airfoil test case has been continuously investigated due to the fact that its geometry contains some of the aerodynamic mechanisms found in turbomachinery applications, but remains computationally simple enough to allow parametric studies \citep{Jacob2005}. In such a configuration, a cylindrical rod is installed upstream of an airfoil, and when introduced in a flow field, the upstream rod generates a turbulent wake (a von \emph{K\'arm\'an} vortex street, consisting of counter-rotating vortices resulting in a nearly constant Strouhal number $St=f\cdot d/U \approx 0.2$, where $f$ is the shedding frequency, $d$ the rod diameter and $U$ the flow velocity), which then interacts with an airfoil downstream. This configuration has been studied by many researchers. \cite{Casalino2003} was the first to investigate this problem using unsteady RANS simulations. The simulations were two-dimensional and the three dimensional effects on noise were modelled using a statistical model coupled with the Ffowcs Williams and Hawkings (FW-H) equation \citep{Ffowcs1969} based on the Lighthill's acoustic analogy. \citet{Boudet2005} reported the first LES computations for this benchmark problem using a finite-volume, compressible LES on
multi-block structured grids. Far-field noise was obtained by coupling the near-field data with a permeable FW-H solver. \citet{Jacob2005} used the experimental acoustic results measured on a rod-airfoil setup for the verification of the numerical broadband noise calculations. Their work has become a standard against which other researchers have benchmarked their code capability and accuracy.

In order to be able to predict sound sources and in particular broadband sources directly, highly accurate transient computational fluid dynamics (CFD) solutions must be obtained. A variety of numerical approaches, including unsteady Reynolds averaged Navier-Stokes (URANS) \citep{Casalino2003, Jacob2005} and large eddy simulations (LES) \citep{Berland2010,Giret2012,Agrawal2016,Jiang2015} have been used to study the rod-airfoil configuration. It should also be noted that the highly unstable phenomena resulting due to the impingement of the turbulent wake on the airfoil, the main source of radiation, are difficult for RANS to resolve. However, it is possible to simulate the flow around the rod-airfoil configuration with the direct numerical simulation (DNS) without any form of turbulence modelled artificially at the cost extraordinary CPU time. The computational effort of three-dimensional DNS scales with the Reynolds number $Re^4$, making the approach not viable for practical airfoil simulations. On the other hand, the computational cost and the available resolution of LES stands between RANS and DNS. LES resolves the large scales eddies (more than 80\% of the turbulent kinetic energy should be resolved \citep{Zhiyin2015}), while the small scale (sub-grid scale (SGS)) are modelled. To address the small turbulent scales, LES requires higher grid resolution and corresponding smaller time-steps than RANS. This results in higher computational costs. Moreover, at rigid boundaries, LES does not respond well. An alternative to compromise the computational costs and the accuracy is the detached eddy simulation (DES). DES is a hybrid model that acts as RANS in the near-wall regions and LES in separate flow areas, integrating advantages of both approaches and being less demanding than the pure LES \citep{Geyer2018}. DES solutions should approach the quality of an LES prediction at optimised processor costs and are therefore a great choice for rod-airfoil simulations. The DES methodology used by \citet{Greschner2004} suggests the efficiency and accuracy of the system for the estimation of the flow over blunt bodies. In a delayed detached eddy simulation (DDES) model, \citet{Zhou2017} used the rod-airfoil configuration to optimise the NACA~0012 airfoil shape with regard to minimum turbulence-interaction noise. The recent work on the rod-airfoil configuration with DES and LES is summarised in Table~\ref{tab:table1}.

\begin{table*}[tb!]
\small\sf\centering
\caption{\label{tab:table1} Numerical studies}
\centering
\begin{tabular}{llll}
\hline
Name&Method & Mesh\\
\hline
\citet{Zhou2017}& DDES& 3D Unstructured\\
\citet{Jiang2015}& LES& 3D Structured\\
\citet{Agrawal2016}& LES& 3D Unstructured\\
\citet{Giret2012}& LES& 3D Unstructured\\
\citet{Galdeano2010}& DES& 3D Unstructured\\
\citet{Berland2010}& LES& 3D Structured\\
\citet{Greschner2008}& DES& 3D Structured\\
\citet{Caraeni2007}& DES& 3D Unstructured\\
\citet{Gerolymos2007}& DES& 3D Structured\\
\citet{Greschner2004}& DES& 3D Unstructured\\
Present study& DDES& 2D and 3D Unstructured\\
\hline
\end{tabular}\\[9pt]
\end{table*}

\noindent
The aforementioned studies were mostly conducted on one rod–airfoil configuration with fixed rod diameter and/or streamwise gap. In a prior experimental study, \cite{Giesler2009} examined the effect of the rod diameter and the streamwise gap on the broadband noise. The present work follows \cite{Giesler2009} by further investigating the effects of these two parameters on the radiated noise, especially the influence of the gap/diameter ratio. The main objectives of the present research are

\begin{enumerate}
    \item Investigation of the influence of geometric parameters (namely the rod diameter and the streamwise gap between rod and airfoil) at low Mach number flow speeds on the noise generation.
    \item Applying the compressible DDES approach to calculate the near-field CFD results and using the Ffowcs-Williams and Hawkings (FWH) approach to predict far-field noise.
\end{enumerate}

\noindent
This paper is organised as follows. First, the experimental setup is presented, followed by the discussion of numerical methodologies. To validate the numerical methodologies, near-field CFD results are compared with the LES results. Finally, the aeroaocustic results are discussed and conclusions are drawn in the last section.

\section{Experimental Setup}
\begin{figure}
\centering
\small
\begin{tikzpicture}[scale=4.4,>=latex]
	\draw[thick,orange,fill,fill opacity=0.4] (-0.41,-0.06) rectangle (-0.31,0.06);
	\draw[thick,orange,fill,fill opacity=0.4] (-0.535,-0.06) rectangle (-0.525,0.06);
	\draw[orange](-0.36,0.06) -- +(65:0.35) node [above, orange] {rod and airfoil at $z=0.72$~m};
	\draw[orange](-0.53,0.06)--+(45:0.45);
	\draw[draw,blue,pattern=north west lines, pattern color=blue, opacity=0.8] (-0.41,-0.05) rectangle (-0.235,0.05);
	\draw[blue](-0.30,-0.05) -- +(280:0.26) node [below, blue] {beamforming integration sector};
	\begin{scope}[scale=0.2,xshift=-2cm]
		\draw[very thick, green](-3,2) node [above right,text width=2cm] {nozzle} .. controls (-2,2) and (-2.0,0.4) .. (-1.0,0.32);
		\draw[very thick, green](-3,-2) .. controls (-2,-2) and (-2.0,-0.4) .. (-1.0,-0.32);
		\draw[->](-2.5,0.2)--(-1.2,0.2);
		\draw[->](-2.5,0)--(-1.2,0) node [at start,left]{};
		\draw[->](-2.5,-0.2)--(-1.2,-0.2);
	\end{scope}
	\draw[red,only marks,mark=*,mark options={scale=0.2}] plot file {array_38.coo};
	\draw[red](0.3,-0.33) node [above right,text width=2.5cm] {microphones\\at $z=0$};
	\draw[-,very thick,gray,opacity=0.4](-0.6,0.065)--(-0.12,0.065);
	\draw[-,very thick,gray,opacity=0.4](-0.6,-0.065)--(-0.12,-0.065);
	\draw[gray](-0.45,-0.065) -- +(-115:0.15) node [below,gray] {side wall};
	\draw[->] (-1.0,-0.42) -- (-1.0,0.6) node [left] {$y$ in m};
	\draw[->] (-1.0,-0.42) -- (0.6,-0.42) node [below] {$x$ in m};
	\foreach \x in {-0.53,-0.41,-0.31,0}
		\draw (\x,-0.40) -- +(0,-0.04) node [pos=1.9,below,rotate=45] {\x};
	\foreach \y in {-0.4,-0.06,0,0.06,0.4} 
		\draw (-0.98,\y) -- +(-0.04,0) node [left] {\y};		
\end{tikzpicture}
\caption{Schematic of the experimental setup inside the aero\-acoustic wind tunnel (top view)}
\label{fig:schematic}
\end{figure}
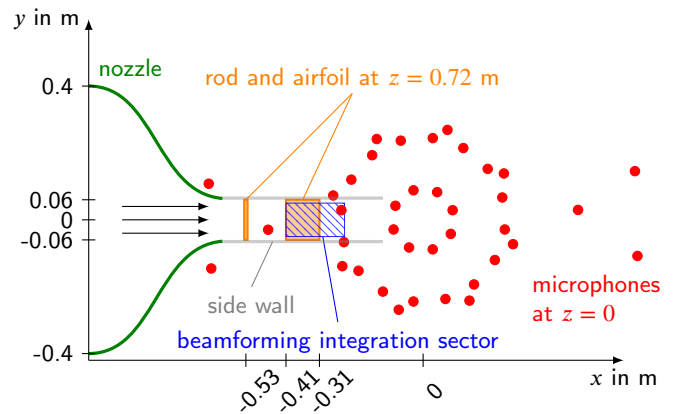

Measurements of the noise generated by the rod-airfoil configuration were conducted in the small aeroacoustic open jet wind tunnel at the Brandenburg University of Technology Cottbus - Senftenberg (see \cite{Sarradj2009}), using rods with diameters of 5~mm, 7~mm, 10~mm, 13~mm and 16~mm and two NACA-type airfoils. The first is a NACA~0012 airfoil and the second a NACA~0018 airfoil, both with a chord length of 100~mm and a span width of 120~mm. The nozzle used in the experiments has a rectangular exit area with dimensions of 120~mm~\texttimes~147~mm. The turbulence intensity in front of the nozzle is below 0.1~\% at a flow speed of 72~m/s. Both the rod and the airfoil model were mounted between side walls, which were covered with an absorbing foam with a thickness of 50~mm in order to reduce potential noise from the wall junction. A schematic of the setup is shown in Fig.~\ref{fig:schematic}. In the current study, only selected experimental results are presented. Additional results can be found in \cite{Giesler2009,Giesler2011}.

\begin{figure*}
\captionsetup[subfigure]{justification=centering}
\captionsetup{singlelinecheck = false, format= hang, justification=raggedright, font=footnotesize, labelsep=space}
        \centering
        \begin{subfigure}[b]{1\columnwidth}
            \centering
            \includegraphics[width=9.8cm]{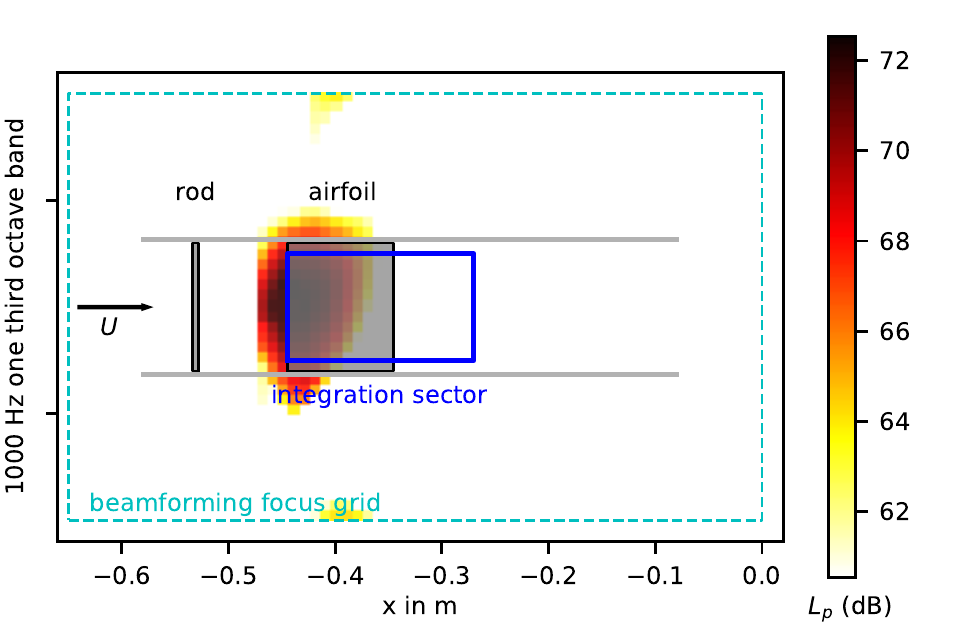}
            \caption[]{}
            \label{fig:soundmap}
        \end{subfigure}
        \hfill
        \begin{subfigure}[b]{1\columnwidth}   
            \centering 
          \includegraphics[width=6.5cm]{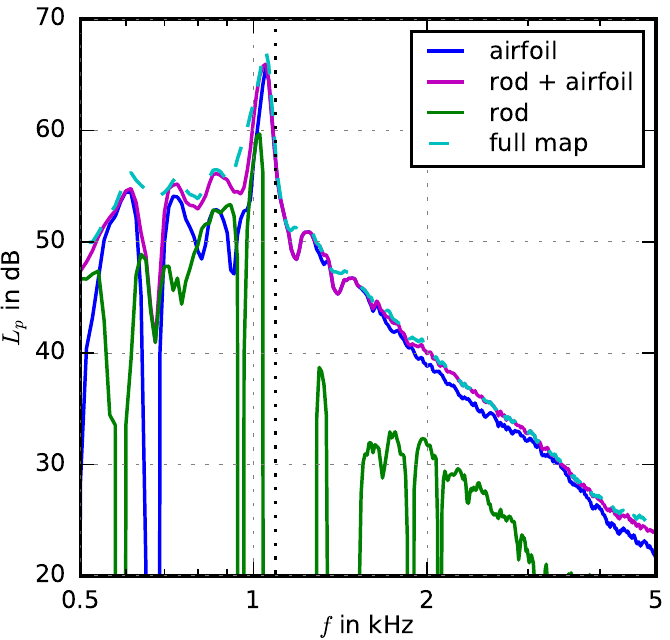}
            \caption[]{}    
            \label{fig:sectors}
        \end{subfigure}
        \caption{Example sound map (left) and sound pressure level spectra obtained for different integration sectors (right) calculated for the configuration with a rod diameter $d$~=~5~mm, a gap width $g$~=~86~mm and a flow velocity of 0.076 ($Re_d = 9.15 \times 10^3$, $Re_c = 1.83 \times 10^5$). (a) DAMAS sound map obtained for the one third octave band with a center frequency of 1~kHz (flow from left to right, beamforming focus grid and chosen integration sector are highlighted), (b) Sound pressure level spectra ($re$~20~$\mu$Pa, frequency resolution 12.5~Hz) obtained by integrating over various integration sectors (dotted line marks cylinder vortex shedding frequency calculated using a Strouhal number of 0.21)}
        \label{fig:acproc}
\end{figure*} 

\noindent
The acoustic measurements were performed with a planar microphone array, consisting of 38 omnidirectional 1/4th inch electret microphone capsules, which have a frequency range from 20~Hz to 20~kHz. They were flush-mounted into an aluminium plate with dimensions of 1.5~m~\texttimes~1.5~m in the layout shown in Fig.~\ref{fig:schematic}. The microphone array was positioned out of the flow in a distance of approximately 0.72~m above the rod and the airfoil. The data were recorded with a sampling frequency of 51.2~kHz and a duration of 40~s using a 24~Bit National Instruments multichannel frontend. In post-processing, which was done using the open source software package \emph{Acoular} from \cite{sarradj2017python}, the data was transferred to the frequency domain by a Fast Fourier Transformation (FFT) on blocks of 4,096 samples, using a Hanning window and an overlap of 50~\% according to the method of \cite{welch1967use}. The resulting spectra were then averaged to yield the cross-spectral matrix and further processed using the DAMAS beamforming algorithm proposed by \cite{Brooks2006}. This algorithm, although computationally expensive, is known for its good performance especially at low frequencies (see, for example, the work of \cite{Herold2017}) and is often used in aeroacoustic studies (see the comparison by \cite{Merino2019}). In the present case, DAMAS was applied to a two-dimensional focus plane parallel to the array, which had an extent of 0.65~m in the streamwise direction and 0.4~m in the spanwise direction. With a high resolution of 0.005~m, this lead to a total of 10,611 grid points. The chosen steering vector was that according to formulation IV in the work of \cite{sarradj2012three}. The result of this beamforming are so-called sound maps, which can be understood as two-dimensional plots of the noise source locations, mapped to the grid points of the beamforming focus grid. As an example, Fig.~\ref{fig:soundmap} shows such a sound map for one of the experimental cases. Also highlighted in this sound map is the total extent of the beamforming focus grid chosen for this study.

In order to obtain quantitative spectra of the far-field noise generated by the airfoil interacting with the inflow turbulence, an integration sector was defined that contains only the noise contributions due to the interaction of the rod-generated turbulence with the airfoil. It is included both in the schematic (Fig.~\ref{fig:schematic}) as well as in the sample sound map (Fig.~\ref{fig:soundmap}). Background noise sources, such as the rod itself, but also the interaction of the boundary layer along the side walls with the airfoil, were excluded from the integration. Sources due to the interaction of the turbulent boundary layer with the airfoil trailing edge, however, were included. The results were then converted to sound pressure level spectra relative to 20~$\mu$Pa. Finally, 6~dB were subtracted to account for the refraction at the array plate. To illustrate the need for a suitable integration sector, Fig.~\ref{fig:sectors} shows sound pressure level spectra obtained from the integration over various sectors, including a sector that only contains the airfoil (which was the one chosen for the processing of the experimental data), a sector that contains both the airfoil and the cylinder, a sector that contains the cylinder only and a sector that contains the full beamforming focus grid and hence all noise sources. It is visible that the vortex shedding from the cylinder leads to strong tonal noise at a frequency of approximately 1.1~kHz, a component that is also visible in the spectrum obtained from integration over the airfoil sector only. The broadband noise above this tonal peak, however, is mainly generated at the leding edge of the airfoil. Figure~\ref{fig:sectors} also contains a theoretical value of the vortex shedding frequency, obtained using a Strouhal number of 0.21. This value was taken from the work of \citep{Schlichting1997}.

\section{Numerical Setup}
\subsection{Method}
Hybrid RANS/LES methods are becoming increasingly popular to compute unsteady detached flows with reasonable performance in terms of efficiency and accuracy. RANS/LES zonal methods rely on two different models, a RANS model and a subgrid-scale model, which are applied in different domains separated by a sharp or dynamic interface, whereas non-zonal methods assume that the governing set of equations is smoothly transitioning from a RANS behaviour to an LES behaviour, based on criteria updated during the computation. The Detached Eddy Simulation approach (DES) was proposed by \cite{Spalart1997}. The original DES proposed combines the RANS and LES in a non-zonal manner. In DES, the length scale is replaced by a modified length scale, $d_{\rm DES}$, which is the minimum between the distance to the wall and a length proportional to the local grid spacing. It is represented mathematically as

\begin{equation}
    d_{\rm DES} = {\rm min}(d,C_{\rm DES}\Delta),
    \label{eq1}
\end{equation}

\begin{equation}
    \Delta = {\rm max}(\Delta_x,\Delta_y,\Delta_z),
    \label{eq2}
\end{equation}

\noindent
where $d$ is the distance to the wall, $\Delta$ is the local maximum grid spacing and $C_{\rm DES} = 0.65$ is the model constant calibrated for isotropic turbulence \cite{Spalart2009}. The original DES length scale, $\Delta$, can become smaller than the boundary layer thickness, leading to an \say{ambiguous grid density} condition for the original DES and erroneous activation of the LES mode within the attached boundary layer. Therefore, \cite{Spalart2006} proposed an improved DES method known as delayed DES to solve the grid-induced separation (GIS) and modeled stress depletion (MSD) problems. The modified length scale

\begin{equation}
    d_{\rm DES} = d-f_d{\rm max}(0,d-C_{\rm DES}\Delta),
    \label{eq3}
\end{equation}

\noindent
is used for the DDES model, where

\begin{equation}
    f_d=1-\tanh \left ( \left [ 8\frac{\nu_t+\nu}{\sqrt{U_{i,j}U_{i,j}}\kappa^2d^2} \right ]^3 \right ),
\end{equation}

\noindent
with $\nu_t$ being the kinematic eddy viscosity, $\nu$ the molecular viscosity, $U_{i,j}$ the velocity gradient and $\kappa$ the Karman constant. In the present work, the solutions of the unsteady Reynolds Average Navier-Stokes (URANS) is carried out using a one-equation Spalart–Allmaras model \citep{SAmodel} for calculating the turbulent viscosity. Away from the wall, the model is transformed into a Smagorinsky LES Sub-Grid Scale (SGS) model \citep{smagorinsky}. For the current work, SU2 \citep{Economon2015}, an open source suite written in C++ and Python to numerically solve partial differential equations (PDE) is used.  For spatial integration, a second-order finite volume scheme is applied on the unstructured grids in SU2 using a standard edge-based data structure on a dual grid with control volumes constructed
using a median-dual, vertex-based scheme \cite{Molina2017}. A dual time-stepping strategy \citep{Jameson} is implemented to achieve a high-order temporal accuracy. In this method, the unsteady problem is converted, at each physical time step, into a series of steady problems, which can then be solved for steady problems using convergence acceleration techniques. For detailed explanation of spatial and temporal integration, please refer to \citep{Molina2017}.

\subsection{Acoustic Solver}
In the framework of the present work, the near-field aerodynamic sources are post-processed to the far-field acoustic pressure using the FWH equations \citep{Ffowcs1969,Farassat2007}, which can be written in time domain as

\begin{equation}
\begin{aligned}
    p'(\textbf{x},t)= {} & \underbrace{\frac{\partial^2 }{\partial x_i \partial x_j}\int_{V}^{ } \left [ T_{ij} \right ]_{\tau=\tau^{\ast }} \frac{d^3 \mathbf{y}}{\left | \mathbf{x-y} \right |}}_{\substack{\text{free-field noise} \\ \text{(quadrupole source)}}} \\ 
    & \underbrace{-\frac{\partial }{\partial x_i}\oint_{S}^{ }\left [ \rho v_i v_j+
\mathcal{P}_{ij} \right ]_{\tau=\tau^{\ast }} \frac{n_j dS(\mathbf{y})}{4 \pi \left | \mathbf{x-y} \right |}}_{\substack{\text{loading noise} \\ \text{(dipole source)}}} \\ 
& + \underbrace{\frac{\partial }{\partial t} \oint_{S}^{ }\left [ \rho v_j \right ]_{\tau=\tau^{\ast }} \frac{n_j dS(\mathbf{y})}{4 \pi \left | \mathbf{x-y} \right |}}_{\substack{\text{thickness noise} \\ \text{(monopole source)}}},
\end{aligned}
\label{eq:ch4-eq5}
\end{equation}

\begin{figure*}
  \centering \includegraphics{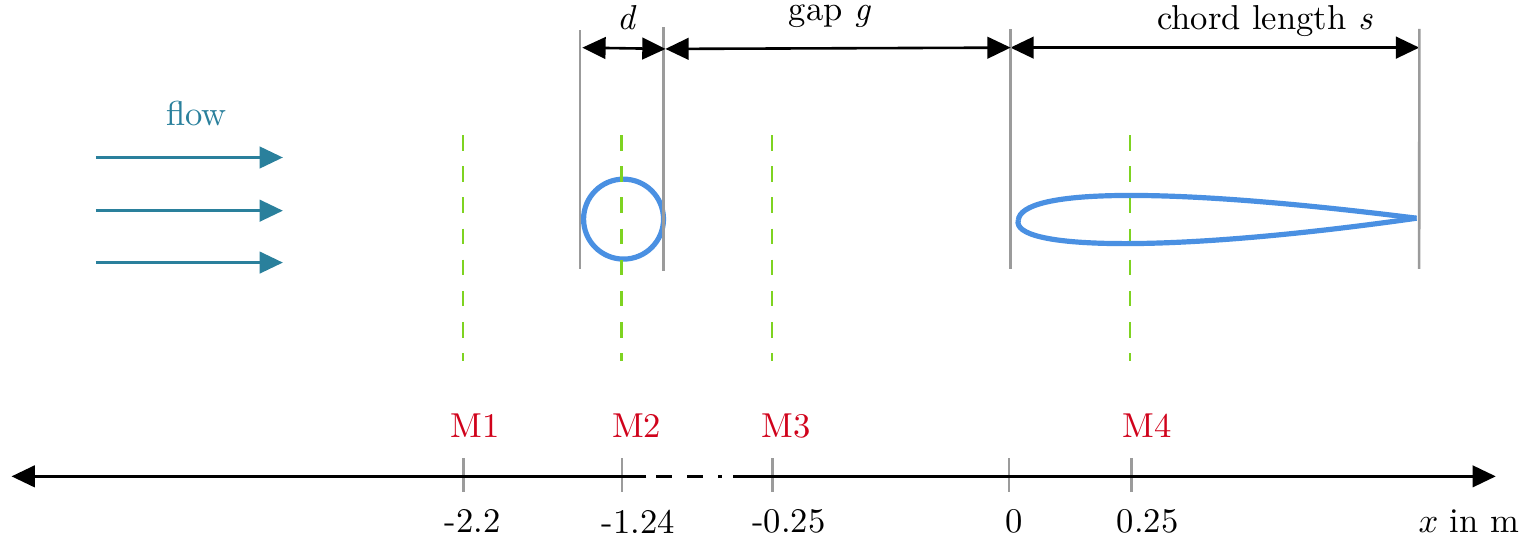}
  \caption{Schematic of the rod-airfoil configuration (side view, the dashed lines denote locations at which velocity profiles were extracted; not to scale)}
  \label{Fig:Fig1}
\end{figure*}

\noindent
where $p'$ is the acoustic pressure, $\mathcal{P}_{ij}$ is the compressive stress tensor, $n_j$ is the surface normal directed into the fluid, and the retarded time is $\tau^\ast = t-(\left | \mathbf{x-y} \right | / c$ with the speed of sound $c$. The quantity $T$ is called Lighthill’s stress tensor. The above equation is the general form, which can then be modified to account for different operating conditions. The first term, the volume integral, is referred to as the quadrupole term. For low Mach number flows $M \ll 1$, the volume source terms of quadrupole order are negligible compared to the surface source terms that are of dipole order \citep{BRENTNER200383}. Hence, the quadrupole term is ignored in the evaluation of Eq.~\eqref{eq:ch4-eq5} for the calculations presented in this paper. The last integral corresponds to a monopole sound field generated by the sound energy flux through the surface $S$. If the surface is rigid or impenetrable (and of course stationary) then this term is zero. Validation of the current FWH implementation is included in \cite{Sharma2020a,Sharma2020}.

\noindent
In the present approach, the flow solver computes the unsteady flow in the computational domain to give the fluctuating flow variables at each time step on the surface points. The acoustic pressure is then calculated in the time-domain by numerically integrating along the FWH surface for each observer location $\mathbf{x}$. In order to compare the broadband predictions based on 2D simulations to the problem of sound radiation from a 3D wing of finite span, additional corrections have to be added to the simulated 2D spectra. \citet{Ewert2009} proposed a simple correction formula as

\begin{equation}
    S_{pp|3D}=\frac{C}{2 \pi} \frac{L}{R} M S_{pp|2D},
    \label{eq:ch4-eq14}
\end{equation}

\noindent
where $C \simeq 2.1$ \citep{Amiet1976}, $L$ is the span width and $R$ is the 2D polar radius vector in the midspan $xy$-plane. It basically denotes the distance to the observer located in the xy-plane at $z = 0$, i.e. $R=\sqrt{x^2+y^2}$.

\subsection{Geometrical Description}
As described above, the rod-airfoil-configuration is a well known and well described problem in aeroacoustics. In a comprehensive analysis published by \citet{Zdravkovich1997}, the significance of the study of this configuration is well established. The first experimental analysis of this problem by \citet{Jacob2005} was performed in order to develop an academic benchmark to test state of the art CAA/CFD techniques. In Fig.~\ref{Fig:Fig1}, the schematic diagram of the configuration to be analysed can be seen, which is consistent with the experimental setup described before. A symmetric NACA~0012 airfoil with a chord length $s$ of 100~mm is located downstream of a rod, both extending 120~mm in the spanwise $z$-direction. The origin of the Cartesian coordinate system is located at the mid-span position of the airfoil leading edge. Two cylindrical rods with diameters $d$ of 5~mm and 16~mm and lengths of 120~mm are used to generate turbulence. The streamwise gap between rod and airfoil, $g$, was set to 86~mm and 124~mm. Simulations are performed at M=0.076 and M=0.21. Corresponding Reynolds numbers are listed in Table~\ref{tab:table2}. A total of 8 cases were tested.

\subsubsection{Computational Domain}
\begin{figure}
    \centering
    \includegraphics[width=0.48\textwidth]{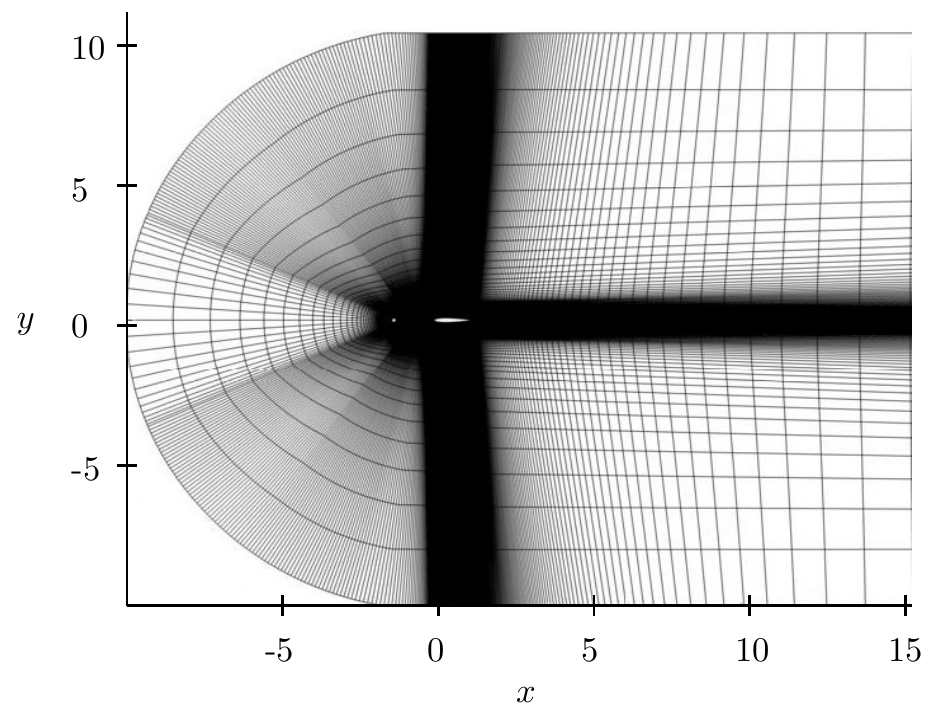}
    \caption{Cross-section view of the computational grid ($x$ and $y$ are given in multiples of the airfoil chord length)}
    \label{fig:Fig_12}
\end{figure}

A two-dimensional computational domain for the configuration was constructed that stretched 20 chord lengths, 20$s$ upstream of the rod, 20$s$ downstream of the airfoil, and 10$s$ to each side side of the rod as shown in Fig.~\ref{fig:Fig_12}. The left side of the boundary was set as a velocity inlet, the upper and the lower boundaries as ambient walls, and the right side as pressure outlet. A fine resolution mesh near the rod and airfoil wall was generated with $y+$~$<$=1.0. A pointwise mesh generation program was used to generate a fully unstructured grid. Corresponding to the experiments and the benchmark, the airfoil was positioned at an angle of attack of zero degrees. To resolve the boundary layer efficiently, an O-type grid and a C-type grid was generated around the rod and the airfoil, respectively. For the mesh elements between the rod and airfoil, the aspect ratio was near unity with an almost constant grid spacing. The resulting computational mesh consists of 1.1 million cells with 300 points around the rod and 400 points around the airfoil in the circumferential direction.

\begin{table*}[htbp!]
\caption{\label{tab:table2} Overview of test cases for numerical studies (given are the rod diameter $d$, the gap width $g$, the flow velocity $U$, the Mach number $M$ and the Reynolds numbers based on rod diameter, $R_d$, airfoil chord length, $R_c$, and Strouhal number based on the rod diameter,  $St_d$)}
\centering
\begin{tabular}{lcccccccc}
\hline
case & $d$ (mm) & $g$ (mm) & U (m/s) & airfoil type & $M$ & $Re_d$ & $Re_c$ & $St_d$\\
\hline
1 & 5 & 86 & 26 & NACA~0012 & 0.076 & $9.15 \times 10^3$ & $1.83 \times 10^5$ & 0.192 \\
2 & 5 & 86 & 72 & NACA~0012 & 0.210 & $2.53 \times 10^4$ & $5.06\times 10^5$ & 0.219 \\
3 & 5 & 124 & 26 & NACA~0012 & 0.076 & $9.15 \times 10^3$ & $1.83 \times 10^5$ & 0.192 \\
4 & 5 & 124 & 72 & NACA~0012 & 0.210 & $2.53 \times 10^4$ & $5.06\times 10^5$ & 0.219 \\ 
5 & 16 & 86 & 26 & NACA~0012 & 0.076 & $2.93 \times 10^4$ & $1.83 \times 10^5$ & 0.195 \\
6 & 16 & 86 & 72 & NACA~0012 & 0.210 & 	$8.11 \times 10^4$ & $5.06\times 10^5$ & 0.222 \\
7 & 16 & 124 & 26 & NACA~0012 & 0.076 & $2.93 \times 10^4$ & $1.83 \times 10^5$ & 0.195 \\
8 & 16 & 124 & 72 & NACA~0012 & 0.210 & $8.11 \times 10^4$ & $5.06\times 10^5$ & 0.222 \\
\hline
\end{tabular}\\[9pt]
\end{table*}

\noindent
In order to test whether the disagreement between the experiment and the simulation can be mainly attributed to the fact that three-dimensional flow structures in the incoming flow cannot be correctly reproduced by a 2D simulation, a three-dimensional simulation was performed for all the cases on a 3D mesh with 100 parallel planes spanning 3 rod diameters. The cell size in between the airfoil and the rod is 0.02~$d$, with $y+$~$<$~1 at the walls and periodic condition applied on the spanwise direction. The calculation was performed in the double-time method using a time step of 0.02~$d/U$ with 40~inner iterations per time step. The simulations were conducted on 200 cores.

\section{Numerical Aerodynamic Results}
\subsection{Flow Features}
\begin{figure*}
    \centering
    \includegraphics[width=0.75\textwidth]{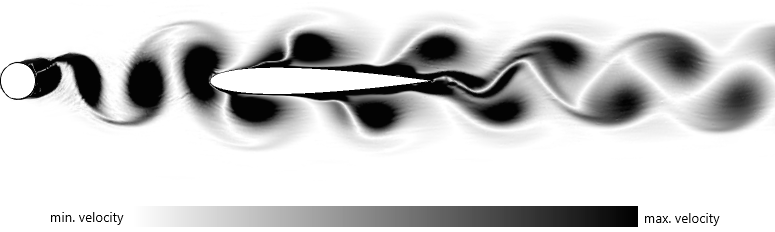}
    \caption{Vortex shedding for case-6 ($d=16$~mm, $g=86$~mm, $M=0.210$, $Re_d=8.11 \times 10^4$, $Re_c=5.06\times 10^5$)}
    \label{fig:Fig_3_1}
\end{figure*}

The first step of the analysis is a qualitative study of the fully developed flow in both 2D and 3D flow simulations. The resulting vortex shedding is captured as periodic oscillations in the wake of the rods. In general, in a two-dimensional simulation a common K\'arm\'an vortex shedding should be observed because the effect of the span is not present to prevent a regular pattern from forming. Fig.~\ref{fig:Fig_3_1} shows the resulting vortex shedding as periodic oscillations in the wake of the rod for case 6 listed in Table~\ref{tab:table2} ($d=16$~mm, $g=86$~mm, $M=0.210$). The non-dimensional shedding frequency, based on the rod diameter, derived from the DDES simulation is $St_d$=0.23$\pm0.02$, which is slightly higher than the value of 0.21 found in the literature (see, for example, the work of \cite{Schlichting1997}). \citet{Greschner2004} explained that the location of the separation point is not determined precisely due to the simple turbulence models in DES, a fact that results in an overpredicted shedding frequency. This could be a potential reason for the disagreement observed in the present study. Fig.~\ref{fig:Fig_3_1} further shows that the vortices shed from the rod are convected downstream towards the leading edge of the airfoil. The larger turbulent structures contained in the wake break down while impinging on the leading edge, whereas the smaller eddies go along the airfoil sides and undergo a distortion.

\begin{figure*}
  \centering
  \small
  \begin{tikzpicture}
  \path (0,0) node {\includegraphics[scale=0.03]{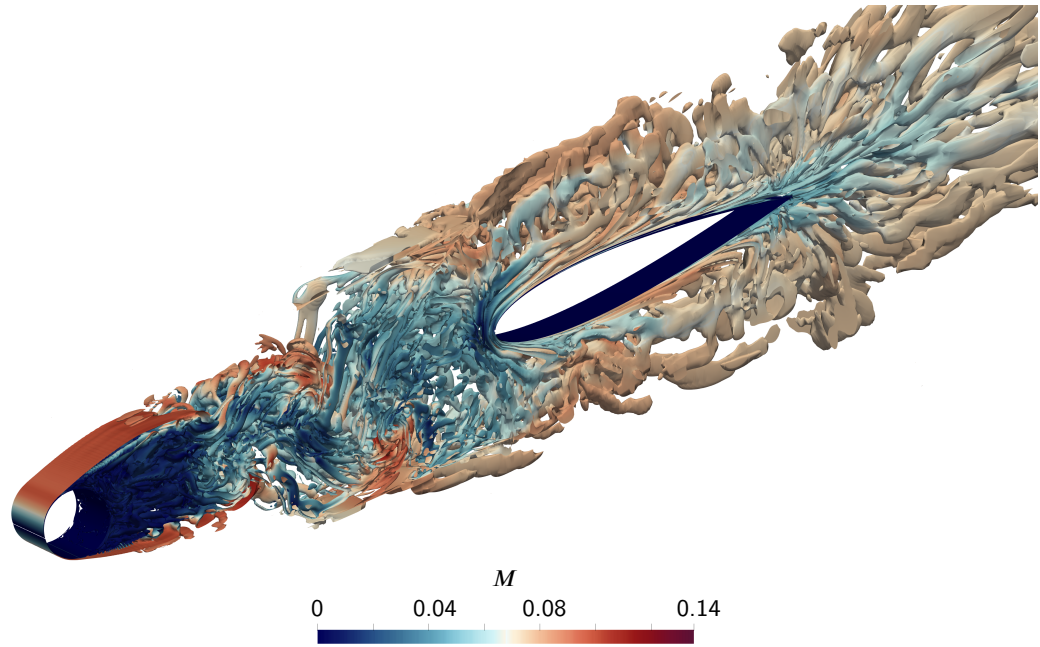}};
  \draw[white,fill] (-3.1,-3.6) rectangle (3.5,-3.0);
  \draw[black](0.5,-3.0) node{$M$};
  \draw[black](-1.97,-3.4) node{0};
  \draw[black](-0.42,-3.4) node{0.04};
  \draw[black](1.02,-3.4) node{0.08};
  \draw[black](3.05,-3.4) node{0.14};
  \end{tikzpicture}
  \caption{Flow structures visualized by iso-surfaces of Mach number-coloured Q-criterion (Q=20) for the case with a rod diameter of 16~mm and a gap of 86~mm at $M$~=~0.076 in the 3D simulation}
  \label{Fig:Fig9}
\end{figure*}

\noindent For the 3D simulation, Fig.~\ref{Fig:Fig9} shows the iso-surfaces of the instantaneous vorticity-field coloured by the Q-criterion. The K\'arm\'an vortex street gradually breaks into small vortices in the wake of the rods, and collides with the leading edge of the airfoil. The dissipation reduces the strength of the vortex with increasing downstream distance.

\subsection{Velocity Profile Comparisons}
To enable comparisons of the mean and root mean square velocity profiles with outcomes described by \cite{Agrawal2016}, 2D and 3D numerical computations were conducted on an additional case ($d$~=~10~mm, $g$~=~100~mm, $s$~=~100~mm, $M$~=~0.2, NACA~0012). The data were sampled for four periods of wake shedding for the numerical simulation. The velocity profiles were extracted at four different streamwise locations, marked as M1 through M4 in the schematic shown in Fig.~\ref{Fig:Fig1}.

\begin{figure*}
\captionsetup[subfigure]{justification=centering}
        \centering
        \begin{subfigure}[b]{1\columnwidth}
            \centering
            \includegraphics[width=\textwidth]{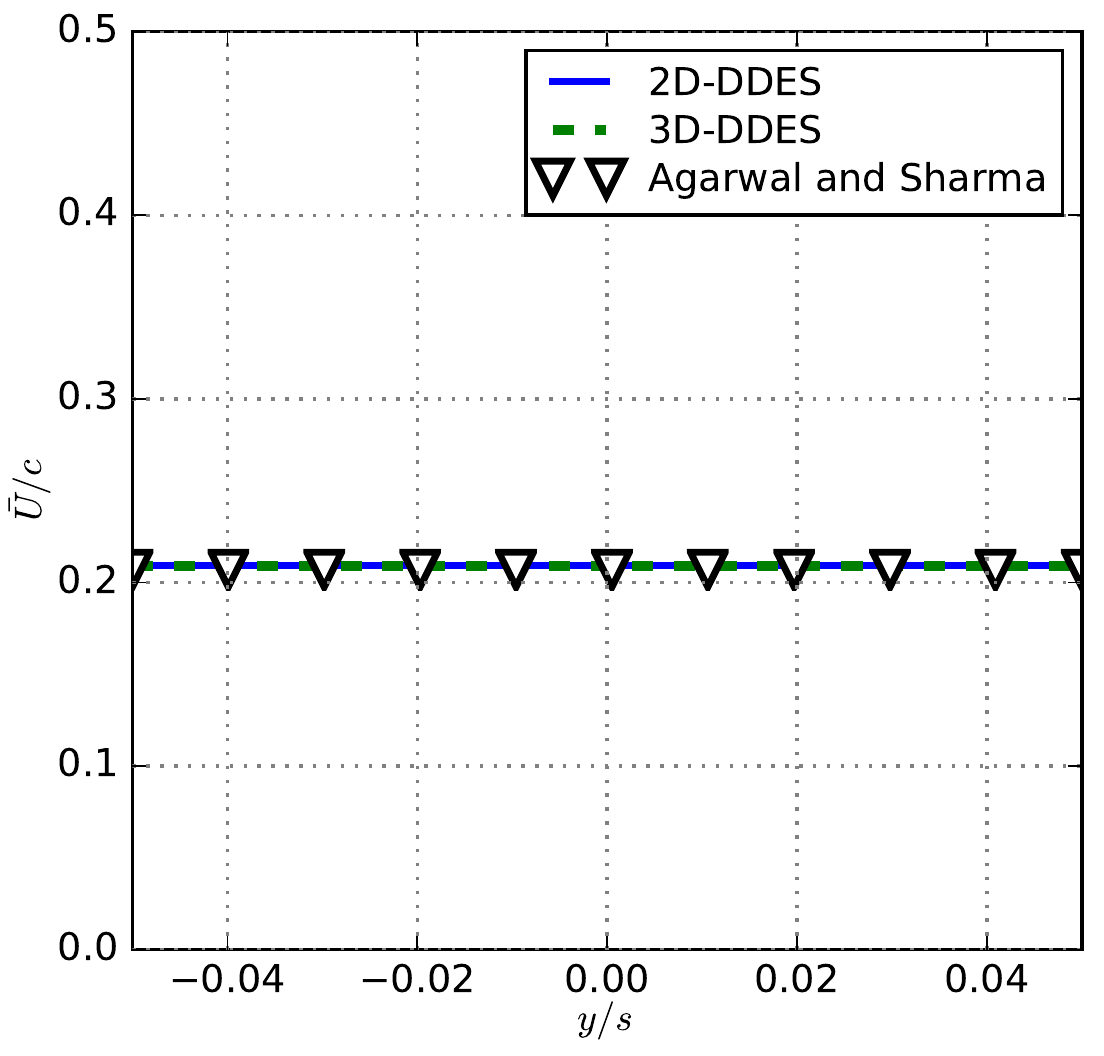}
            \caption[]%
            {} 
            \label{fig:fig3a}
        \end{subfigure}
        \hfill
        \begin{subfigure}[b]{1\columnwidth}   
            \centering 
           \includegraphics[width=\textwidth]{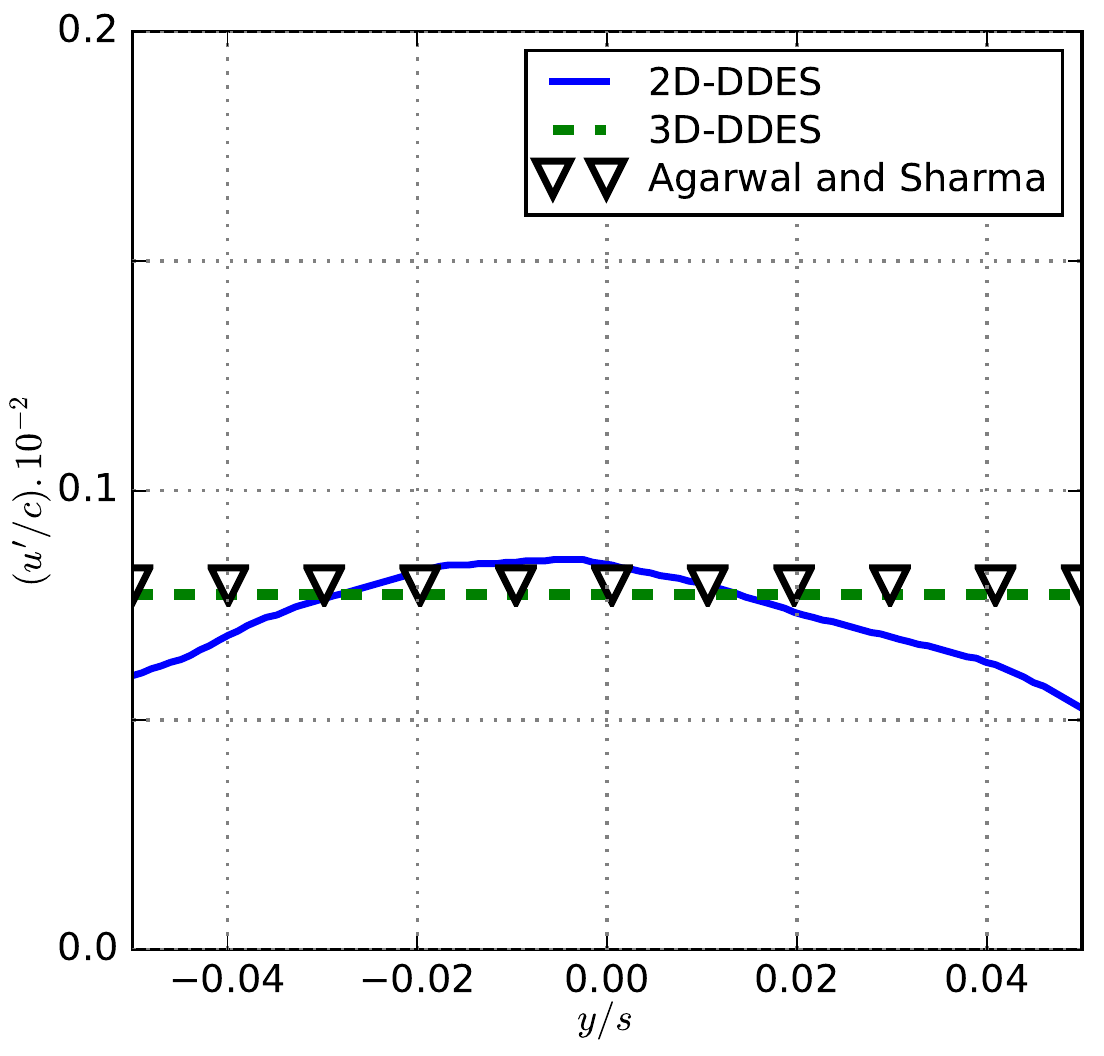}
            \caption[]%
            {}    
            \label{fig:fig3b}
        \end{subfigure}
         \vskip\baselineskip
        \begin{subfigure}[b]{1\columnwidth}  
            \centering 
           \includegraphics[width=\textwidth]{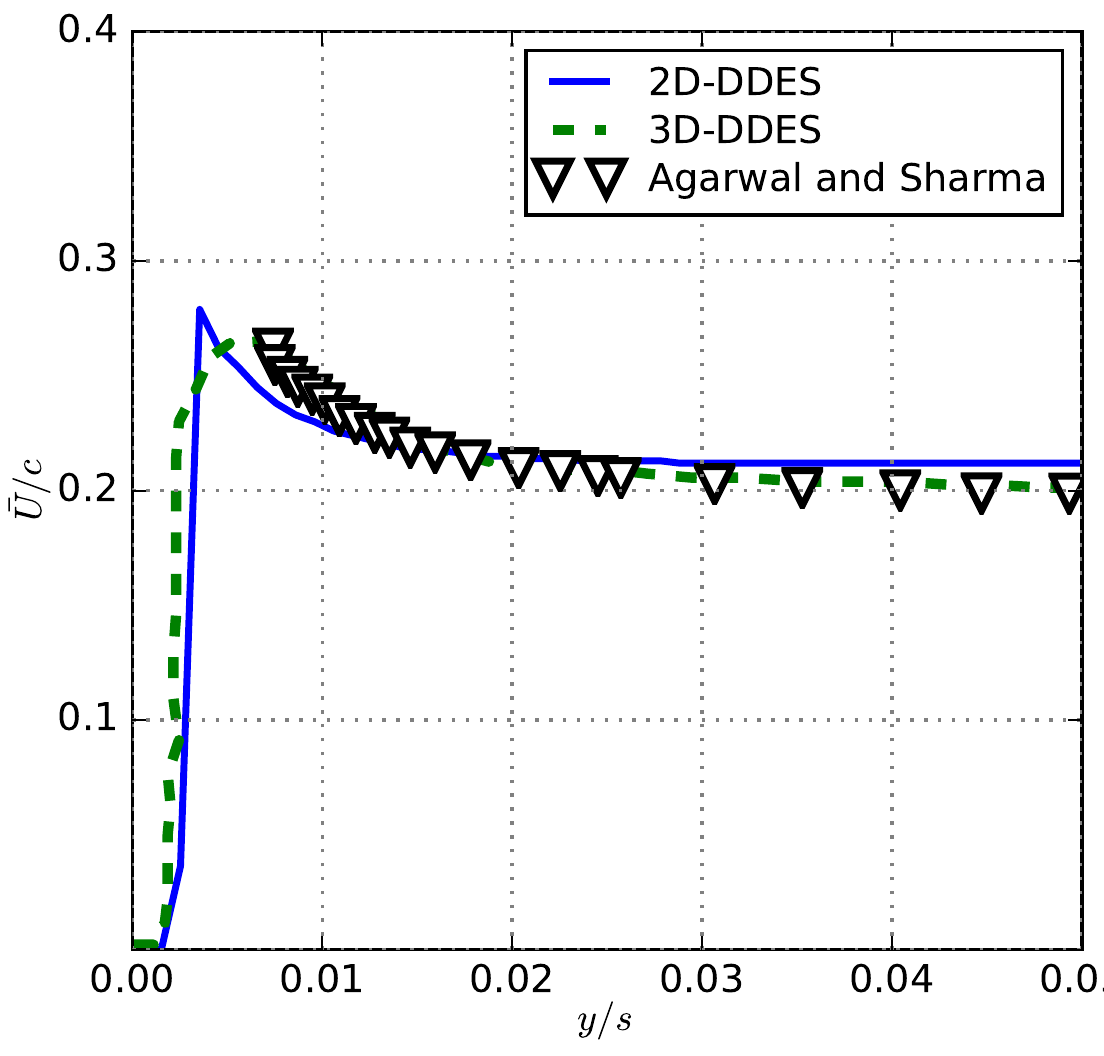}
            \caption[]%
            {}    
            \label{fig:fig3c}
        \end{subfigure}
        \hfill
        \centering
        \begin{subfigure}[b]{1\columnwidth}
            \centering
           \includegraphics[width=\textwidth]{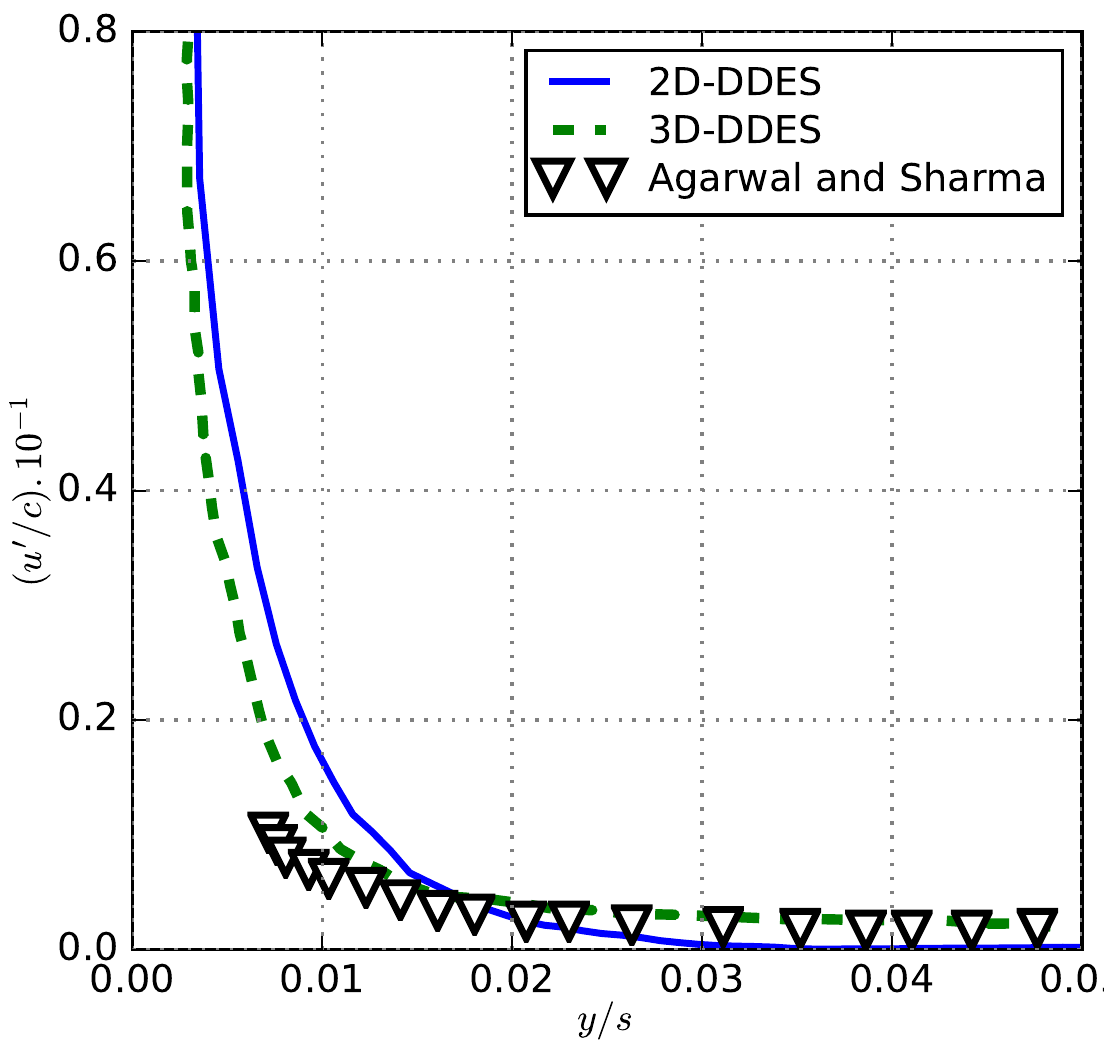}
            \caption[]%
            {} 
            \label{fig:fig3d}
        \end{subfigure}
        \caption{Comparison of velocity profiles obtained at markers M1 and M2 (see Fig.~\ref{Fig:Fig1}) from the simulations with experimental data from \citep{Agrawal2016}. (a) $\overline{U}/c$ at $x/c_l$~=~-2.2 (M1), (b) $u'/c$ at $x/c_l$~=~-2.2 (M1), (c) $\overline{U}/c$ at $x/c_l$~=~-1.24 (M2), (d) $u'/c$ at $x/c_l$~=~-1.24 (M2).}
        \label{fig:fig2}
\end{figure*} 

\begin{figure*}
\captionsetup[subfigure]{justification=centering}
        \centering
        \begin{subfigure}[b]{1\columnwidth}
            \centering
            \includegraphics[width=\textwidth]{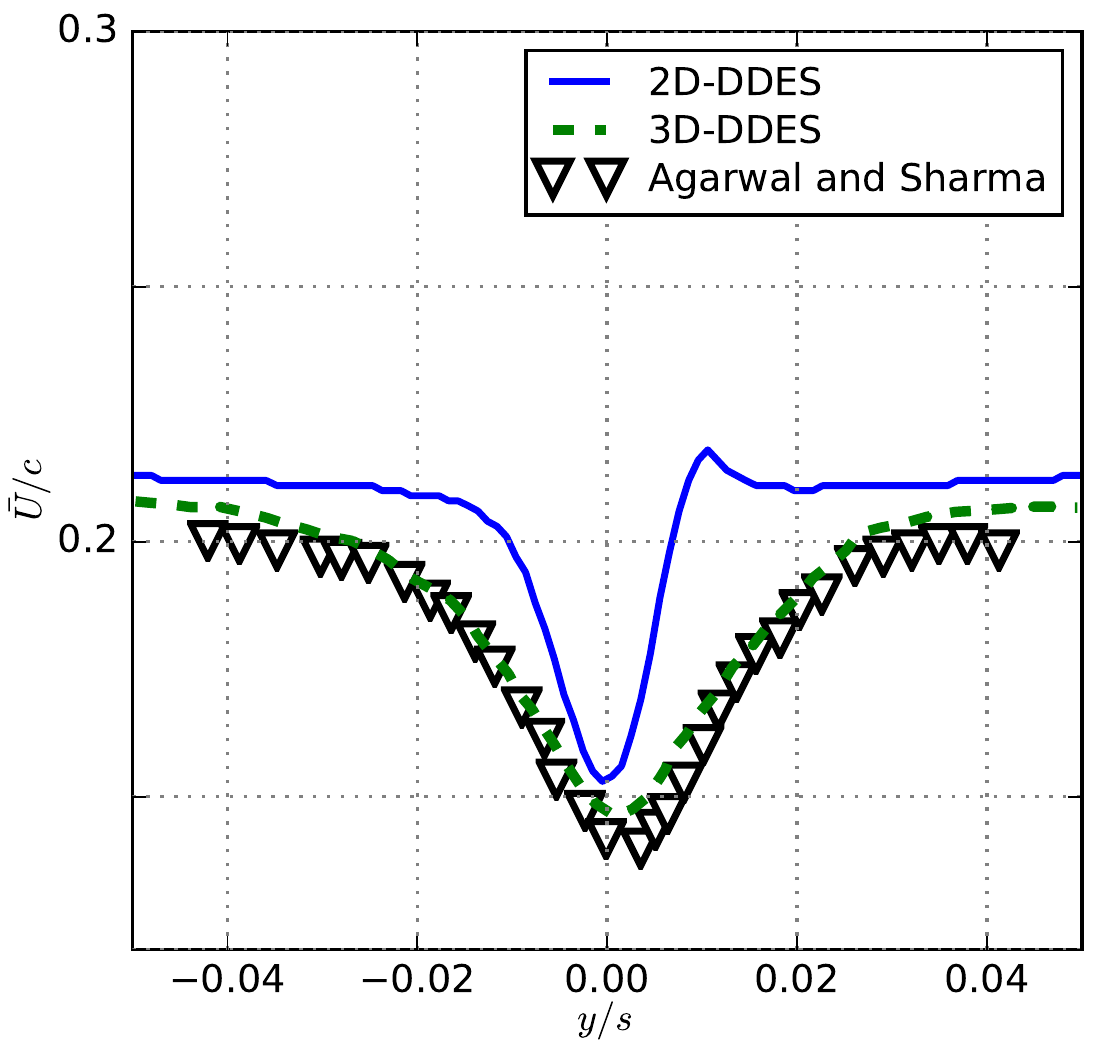}
            \caption[]%
            {} 
            \label{fig:fig4a}
        \end{subfigure}
        \hfill
        \begin{subfigure}[b]{1\columnwidth}   
            \centering 
           \includegraphics[width=\textwidth]{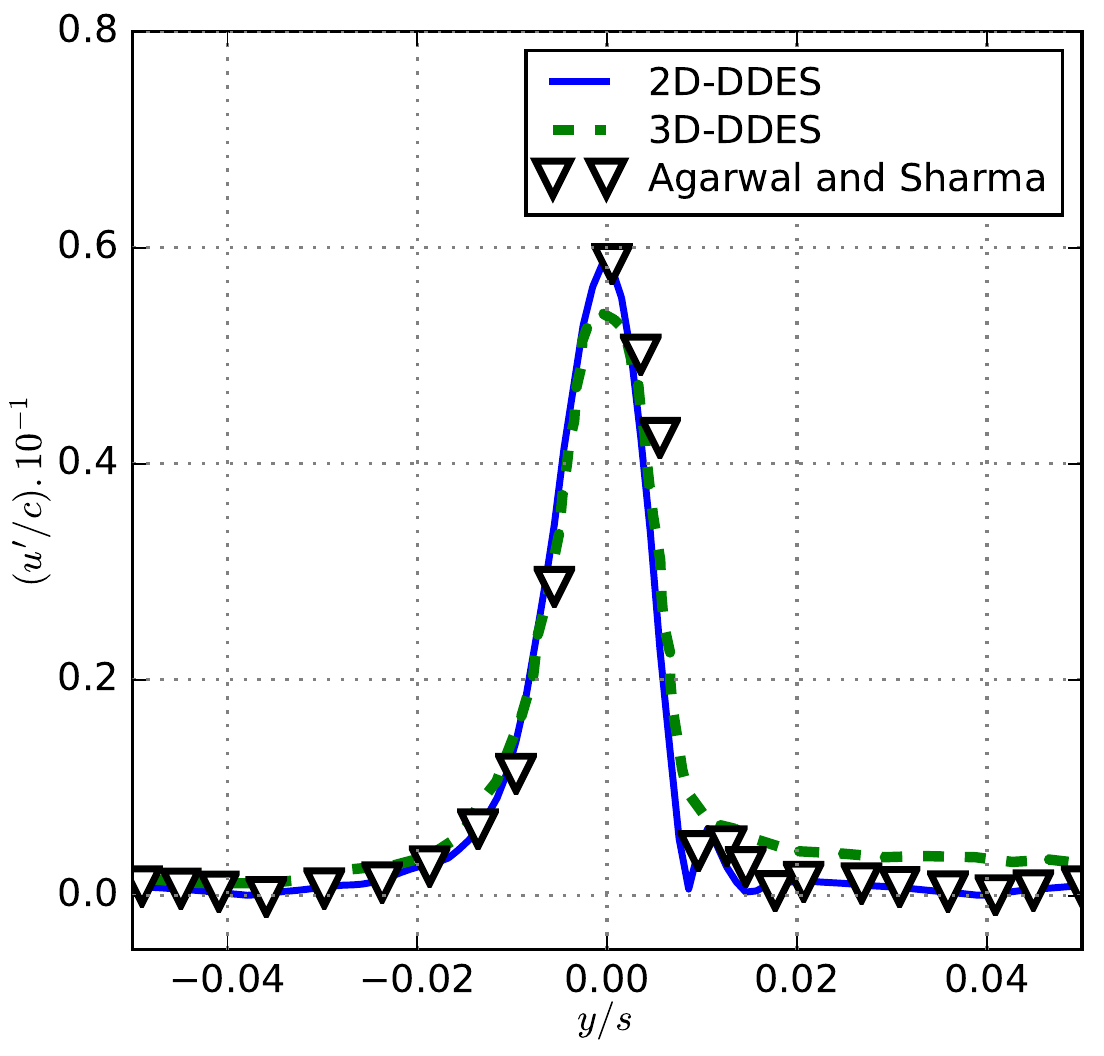}
            \caption[]%
            {}    
            \label{fig:fig4b}
        \end{subfigure}
         \vskip\baselineskip
        \begin{subfigure}[b]{1\columnwidth}  
            \centering 
           \includegraphics[width=\textwidth]{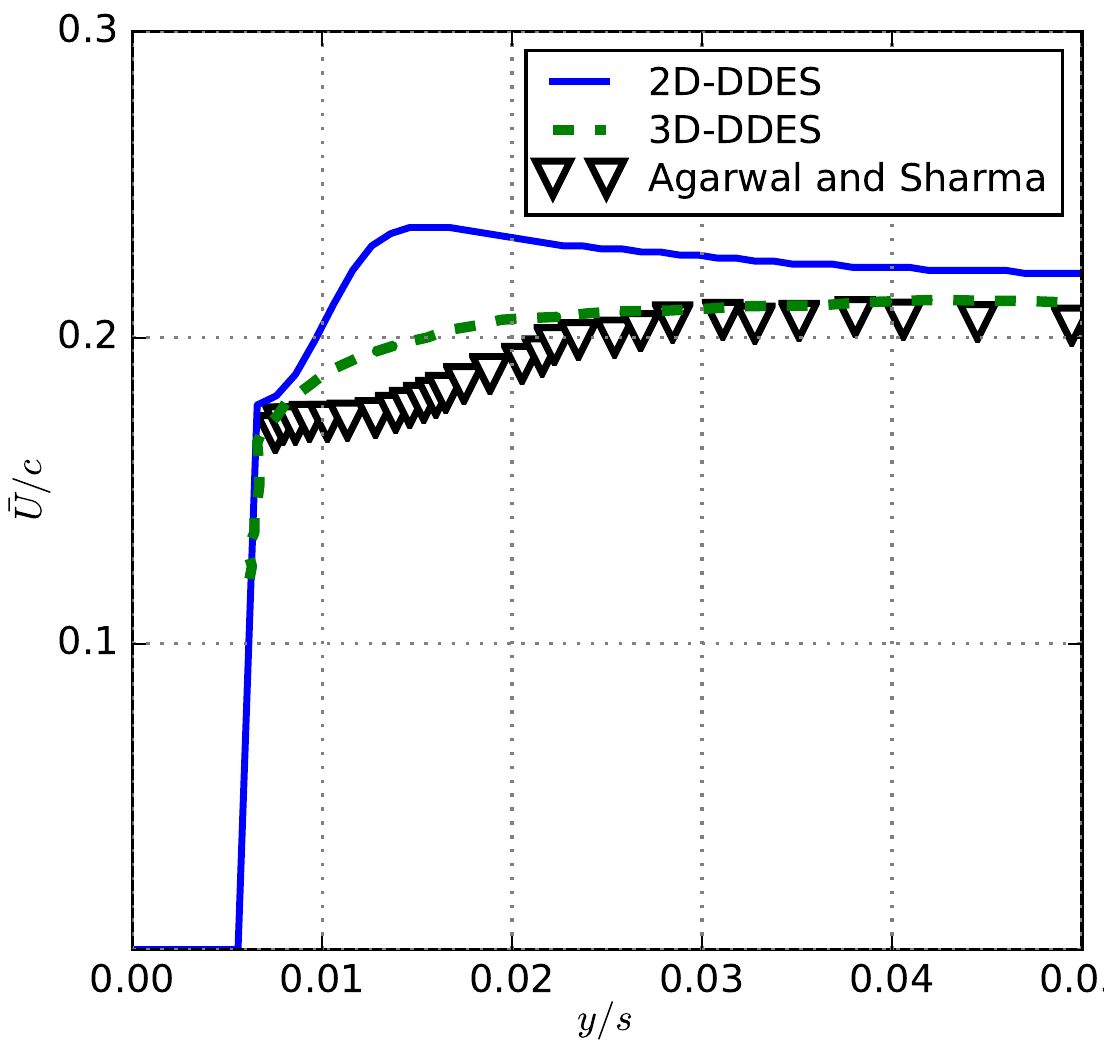}
            \caption[]%
            {}    
            \label{fig:fig4c}
        \end{subfigure}
        \hfill
        \centering
        \begin{subfigure}[b]{1\columnwidth}
            \centering
           \includegraphics[width=\textwidth]{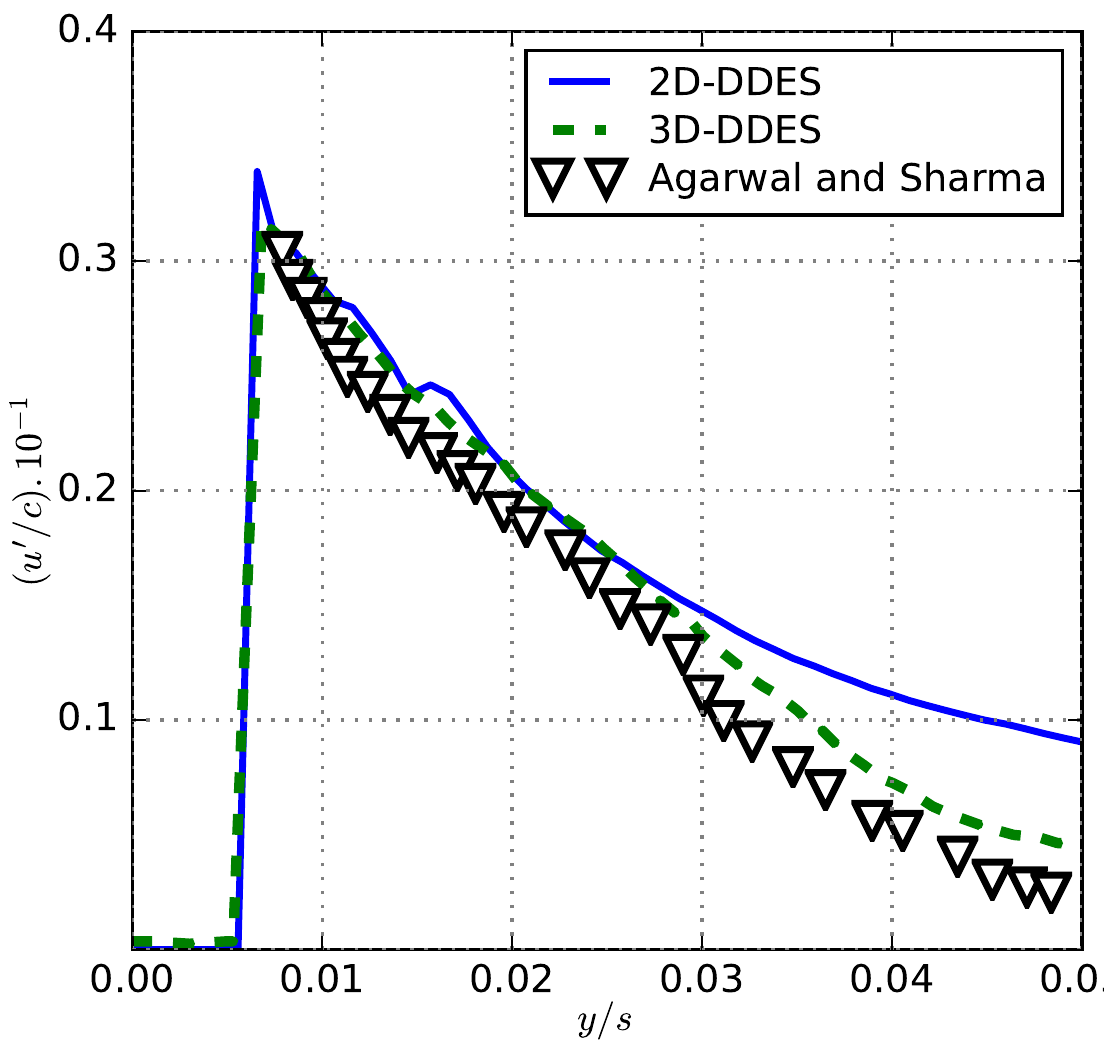}
            \caption[]%
            {} 
            \label{fig:fig4d}
        \end{subfigure}
        \caption{Comparison of velocity profiles obtained at markers M3 and M4 (see Fig.~\ref{Fig:Fig1}) from the simulations with experimental data from \citep{Agrawal2016}. (a) $\overline{U}/c$ at $x/c_l$~=~-0.25 (M3), (b) $u'/c$ at $x/c_l$~=~-0.25 (M3), (c) $\overline{U}/c$ at $x/c_l$~=~0.25 (M4), (d) $u'/c$ at $x/c_l$~=~0.25 (M4).}
        \label{fig:fig3}
\end{figure*}

\noindent The streamwise mean and fluctuating velocities are extracted at the M1 marker at $x/s$~=~-2.2, which is upstream of the rod, and are compared with the experimental work by \cite{Agrawal2016} as shown in Figs.~\ref{fig:fig3a} and \ref{fig:fig3b}. The velocities, basically the conditions of the free stream, are clearly predicted by the DDES. The velocity profiles at M2 ($x/s$~=~-1.24), which bisects the rod into equal segments, are shown in Figs.~\ref{fig:fig3c} and \ref{fig:fig3d}. The profiles derived from the simulations are again in good agreement with the experiments. The velocity profiles at M3 ($x / s$~=-0.25) and M4 ($x / s$~=0.25) are shown in Figs.~\ref{fig:fig4a}-\ref{fig:fig4b} and Figs.~\ref{fig:fig4c}-\ref{fig:fig4d}, respectively. The mean velocity at M3 (upstream of the airfoil) is slightly overpredicted, whereas the rms values of the fluctuating speed are in line with the experimental data at this location. In general, this demonstrates that, as shown earlier by \citet{Greschner2004}, the 3D-DDES findings are consistent with the experimental data. The rod wake turbulence due to its high intensity and length scale determines the velocity profiles on the airfoil rather than the boundary layer development \citep{Sharma2019}. It is not surprising that a purely 2D simulation result differs significantly from measurements. Numerically, the flow does not have a third dimension to go to so the levels of fluctuation tend to be much too strong compared to a 3D simulation and measurements. This is consistent with the fact that there is no 2D turbulence. In DES, away from the wall, the active mode is LES mode where 3D turbulence is supposed to freely develop -- but it can't because there is no third dimension and the effect of the turbulence model is very weak because the SGS drives up the destruction term in the turbulence model. So in that region, 2D simulations can be wrong as reported by \cite{Spalart2009,Mockett2009}.

\section{Aeroacoustic Results}
In the first part of this section, the results from the experiments will be discussed. In the second part, the post-processed data obtained from the CFD solver using the FWH solver will be evaluated and the acoustic results from the simulations will be compared with the experimental ones.

\subsection{Experimental Results}
\label{exp-res}
The experimental results obtained for the configurations summarized in Table~\ref{tab:table2} are shown as 1/3 octave-band sound pressure levels in Fig.~\ref{fig:figc2} and \ref{fig:figc1}. As was already observed in Fig.~\ref{fig:sectors}, each spectrum exhibits a tonal peak at the 1/3 octave-band that contains the vortex shedding tone of the rod. This is due to the fact that the periodic K\'arm\'an vortices shed by the upstream rod interact with the airfoil leading edge. Regarding the broadband noise, notable differences are visible depending on the exact configuration, meaning the cylinder diameter $d$ and the gap width $g$.

\subsubsection{Effect of rod diameter ($d$):}
\label{sec:rod-diameter}
\begin{figure*}
\captionsetup[subfigure]{justification=centering}
        \centering
        \begin{subfigure}[b]{1\columnwidth}
            \centering
            \includegraphics[width=\textwidth]{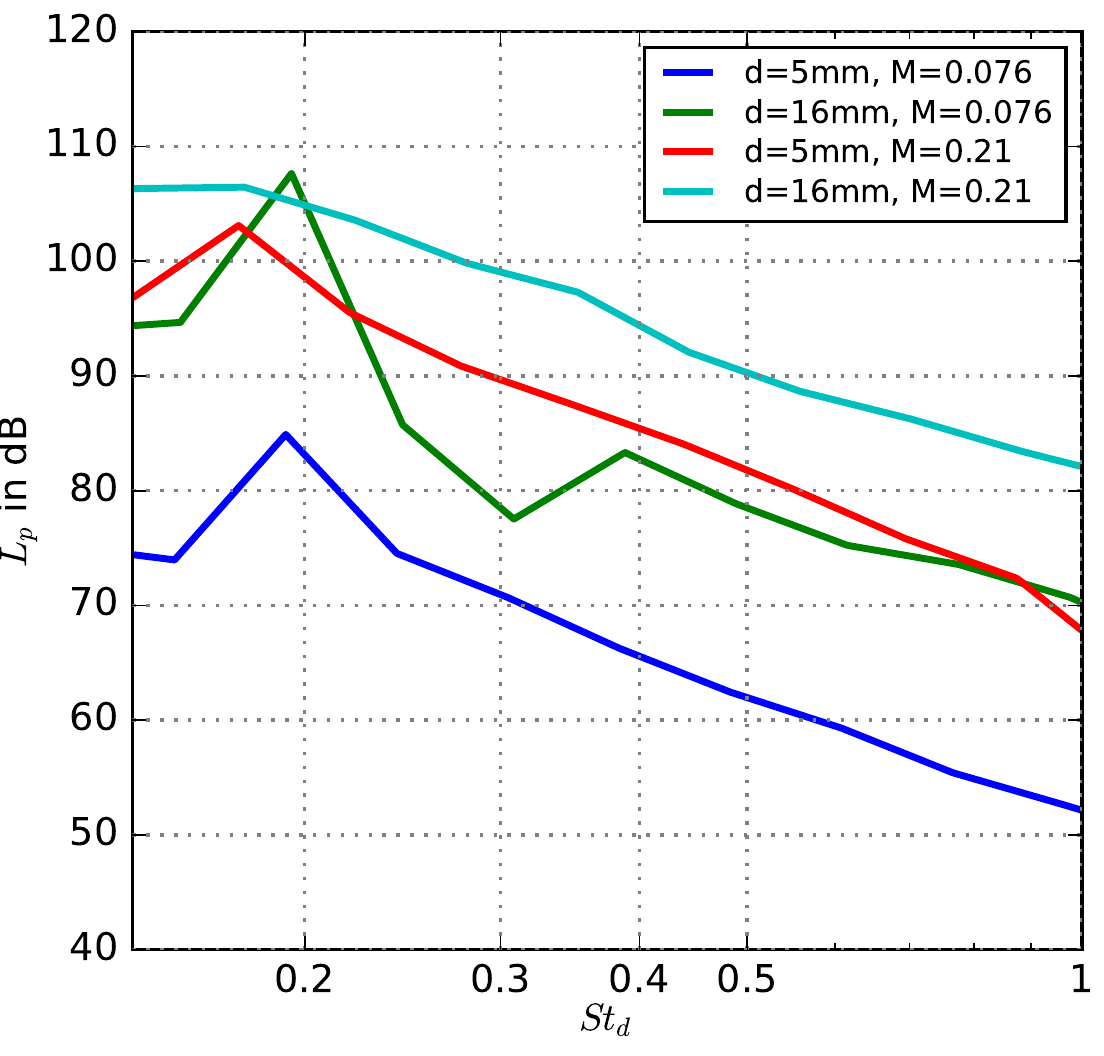}
            \caption[]%
            {} 
            \label{fig:figc2a}
        \end{subfigure}
        \hfill
        \begin{subfigure}[b]{1\columnwidth}   
            \centering 
           \includegraphics[width=\textwidth]{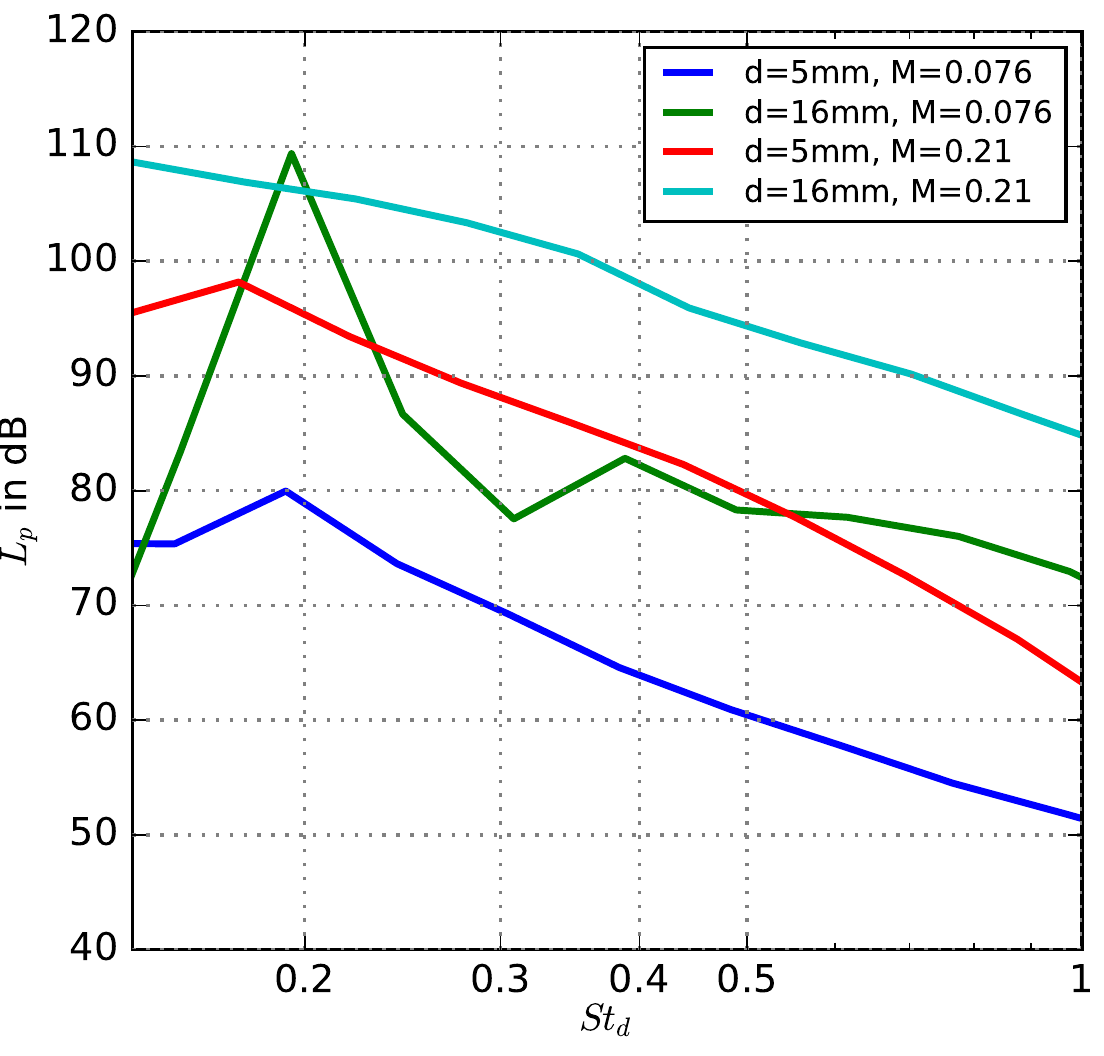}
            \caption[]%
            {}    
            \label{fig:figc2b}
        \end{subfigure}
        \caption[]{Experimental results: 1/3 octave band sound pressure level spectra showing the effect of rod diameter $d$ on the noise generation. (a) g=86\,mm, (b) g=124\,mm.}
        \label{fig:figc2}
\end{figure*}

The sound spectra in Fig.~\ref{fig:figc2} show that the dominant frequency is close to the rod vortex shedding frequency and the corresponding Strouhal number $St_d$ is about 0.2. For the small rod with a diameter of 5~mm with $g = 86$~mm, it can be clearly seen that the peak levels are 85~dB and 103~dB at $M$=0.076 and $M$=0.21, respectively. In the case of the 16~mm diameter rod, the peak levels are 108~dB and 106~dB at $M$~=~0.076 and $M$~=~0.21, respectively.
These results indicate that the intensity of the vortex-structure interaction at the airfoil leading edge decreases with decreasing rod diameter. In some ways this corresponds to the noise reduction when the airfoil thickness is increased \cite{Gershfeld2004b,Giesler2009}. The similar behaviour can be observed with $g = 124$~mm. The observation that the cases with a rod of smaller diameter emit less noise can be justified by the theory of rapid distortion that says that the vortices shed from the smaller rod are less deformed when passing through the leading edge of the airfoil \cite{Li2014}. This results in smaller pressure fluctuations on the surface of the airfoil.

\subsubsection{Effect of streamwise gap ($g$):}
\label{sec:streamwise-gap}
\begin{figure*}
\captionsetup[subfigure]{justification=centering}
        \centering
        \begin{subfigure}[b]{1\columnwidth}
            \centering
            \includegraphics[width=\textwidth]{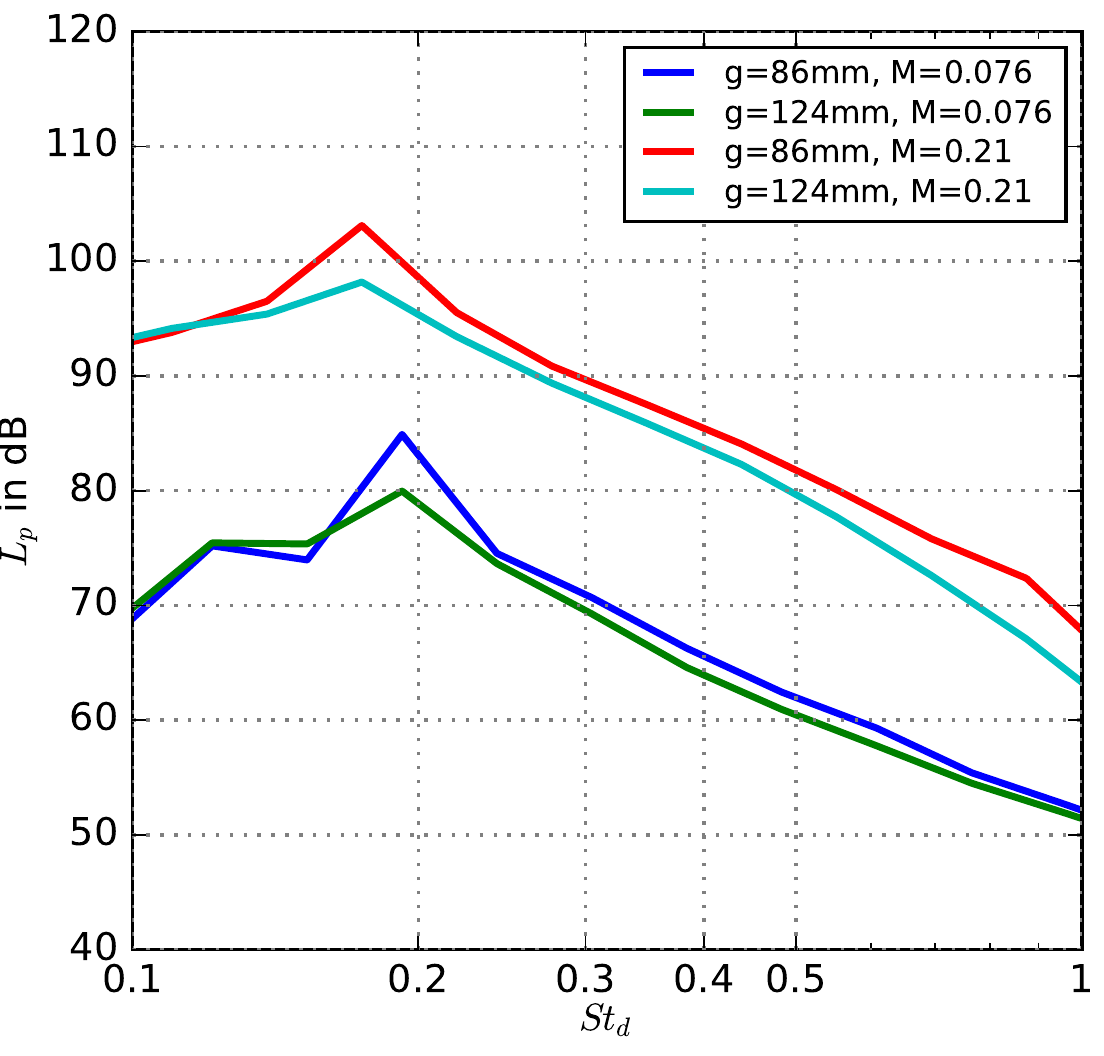}
            \caption[]%
            {} 
            \label{fig:figc1a}
        \end{subfigure}
        \hfill
        \begin{subfigure}[b]{1\columnwidth}   
            \centering 
           \includegraphics[width=\textwidth]{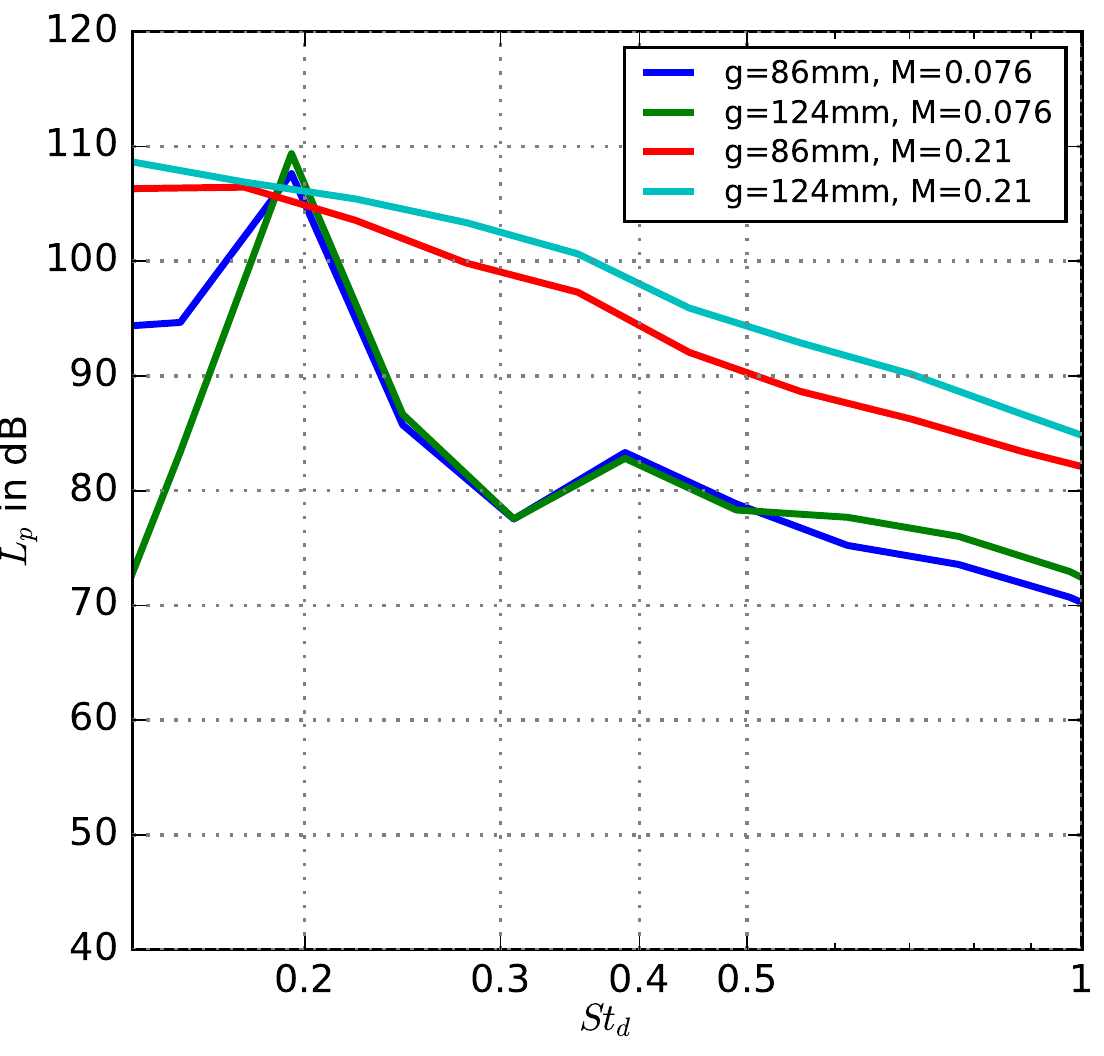}
            \caption[]%
            {}    
            \label{fig:figc1b}
        \end{subfigure}
        \caption[]{Experimental results: 1/3 octave band sound pressure level spectra showing the effect of gap width $g$ on the noise generation. (a) d=5\,mm, (b) d=16\,mm.}
        \label{fig:figc1}
\end{figure*}

When looking at cases with identical rod diameter, but different gap width in Fig. \ref{fig:figc1}, a strange observation can be made: For the thinner rod diameter of 5~mm, the cases with the smaller gap width of 86~mm lead to slightly higher amplitudes, while for the thicker rod diameter of 16~mm, the cases with the larger gap width generate slightly more noise. 

\noindent In order to obtain a better understanding of the basic trends observed in the experimental results, it is reasonable to consider the relation between the noise generated at the leading edge of an airfoil and the properties of the turbulent inflow. When examining the fundamental turbulence interaction noise prediction model by \cite{Amiet1975} it can be seen that the turbulence intensity of the inflow has a strong effect on the amplitude of the resulting noise. The integral length scale of the incoming turbulence, on the other hand, will have a strong impact on the spectral shape of the resulting far field noise, but only a small effect on its amplitude. An increase in integral length scale, and hence an increase in the size of the turbulent eddies, mainly leads to a broadening of the spectral peak and a shift of this peak towards low frequencies (see, for example, \citep{Geyer2018B}). Since the airfoil was the same in all cases examined here, the observed differences between the broadband content of the measured sound pressure level spectra of the various configurations are a direct result of the corresponding turbulence that hits the airfoil leading edge.

\noindent For the cases where the gap width is large compared to the rod diameter, the detailed study by \cite{Roach1987} on the turbulence generated by different kinds of grids may be used to shed further light on the turbulence generation. According to this work, the turbulence intensity generated by a grid (including grids that simply consist of a parallel array of circular cylinders) decays with $(x/d)^{-5/7}$, where $d$ is the rod or wire diameter and $x$ is the streamwise distance from the grid. Regarding the integral length scale of the turbulence, Roach states that it can be approximated as being proportional to $\sqrt{xd}$. Hence, the integral length scale increases with increasing distance. However, these approximations are valid only for sufficiently large distances from the grid, where the turbulence can be assumed to be homogeneous and isotropic. In the present investigation, this is most likely true for the small cylinder diameter, for which the ratio of gap width to cylinder diameter, $g/d$, equals 17.2 for the smaller gap of 86~mm and 24.8 for the larger gap of 124~mm. In both cases, it can be assumed that the turbulence interacting with the leading edge is close to homogeneous. Using the approximations derived by Roach, the cases with the smaller gap width result in a higher turbulence intensity. This, in turn, would lead to a higher sound pressure level according to the model of Amiet, which is exactly what can be seen in the experimental results in Fig.~\ref{fig:figc1a}.

\noindent However, for the thicker rod with a diameter of 16~mm, the present gap widths of 86~mm ($g/d = 5.375$) and 124~mm ($g/d = 7.75$) mean that the leading edge of the airfoil will be quite close to the airfoil. In this region, the simple approximations given by Roach will not be valid. It is rather likely that the large diameter of the cylinder leads to a large distance between the shear layers on both sides and a large separation bubble. For the cases with the smaller gap width of 86~mm (cases 5 and 6), the airfoil leading edge may be located in a region of reduced turbulence inside the or close to the separation bubble, which explains the lower noise generation. This hypothesis agrees well with results obtained for so-called tandem cylinder configurations \citep{lockard2007tandem}. Depending on the distance between the two cylinders, it is possible that the downstream cylinder is located between the shear layers separating from the upstream cylinder (see, for example, \citep{wu1994spanwise,alam2014aerodynamics}).

\subsection{Numerical Results and Comparison with Experiments}
The far-field spectra of the radiated sound is calculated at the observer's location at the center of the planar microphone array (0,0,0) (please refer to Fig.~\ref{fig:schematic}) using the integration method \citep{Ffowcs1969} of Ffowcs Williams and Hawkings with the rigid wall surfaces as the integration surfaces. For low Mach number flows, omitting the quadrupole source has little effect on the noise radiation \citep{BRENTNER200383},  therefore, dipole sources are examined in the present study. This has been achieved for a complete simulation time of 1.0~s out of which the initial 0.2~s were not considered to eliminate the transients. The FWH surface integrals were calculated for both the circular rod and the airfoil as the sources of sound. However, only the airfoil was considered as the source to compare the results with the experiments. Accordingly, in accordance with \cite{Giesler2009,Giesler2011} the study focuses on the high-frequency component of the spectrum with frequencies greater than the frequency of the vortex shedding from the rod.

\begin{figure*}
\captionsetup[subfigure]{justification=centering}
        \centering
        \begin{subfigure}[b]{1\columnwidth}
            \centering
            \includegraphics[width=\textwidth]{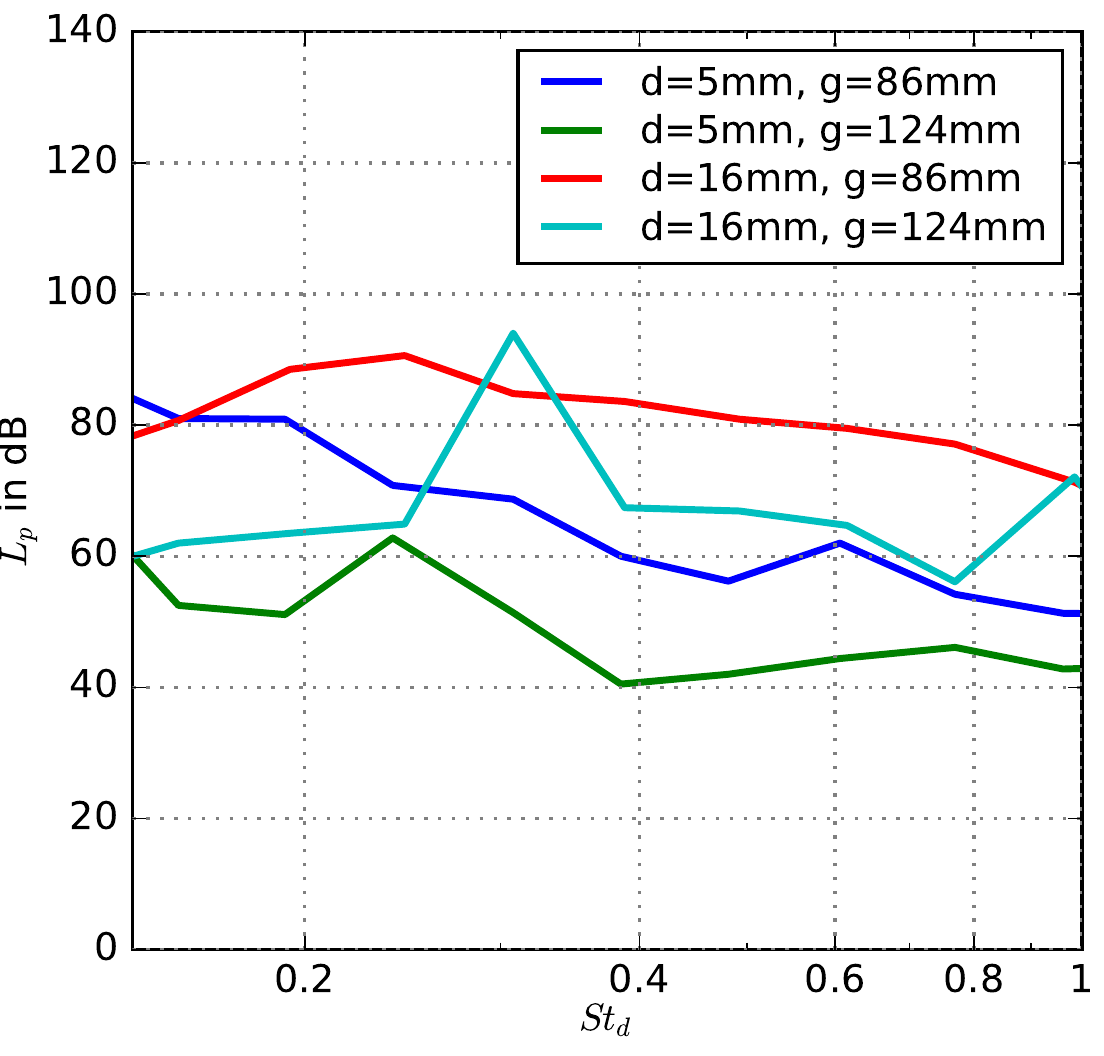}
            \caption[]%
            {} 
            \label{fig:fig5a}
        \end{subfigure}
        \hfill
        \begin{subfigure}[b]{1\columnwidth}   
            \centering 
           \includegraphics[width=\textwidth]{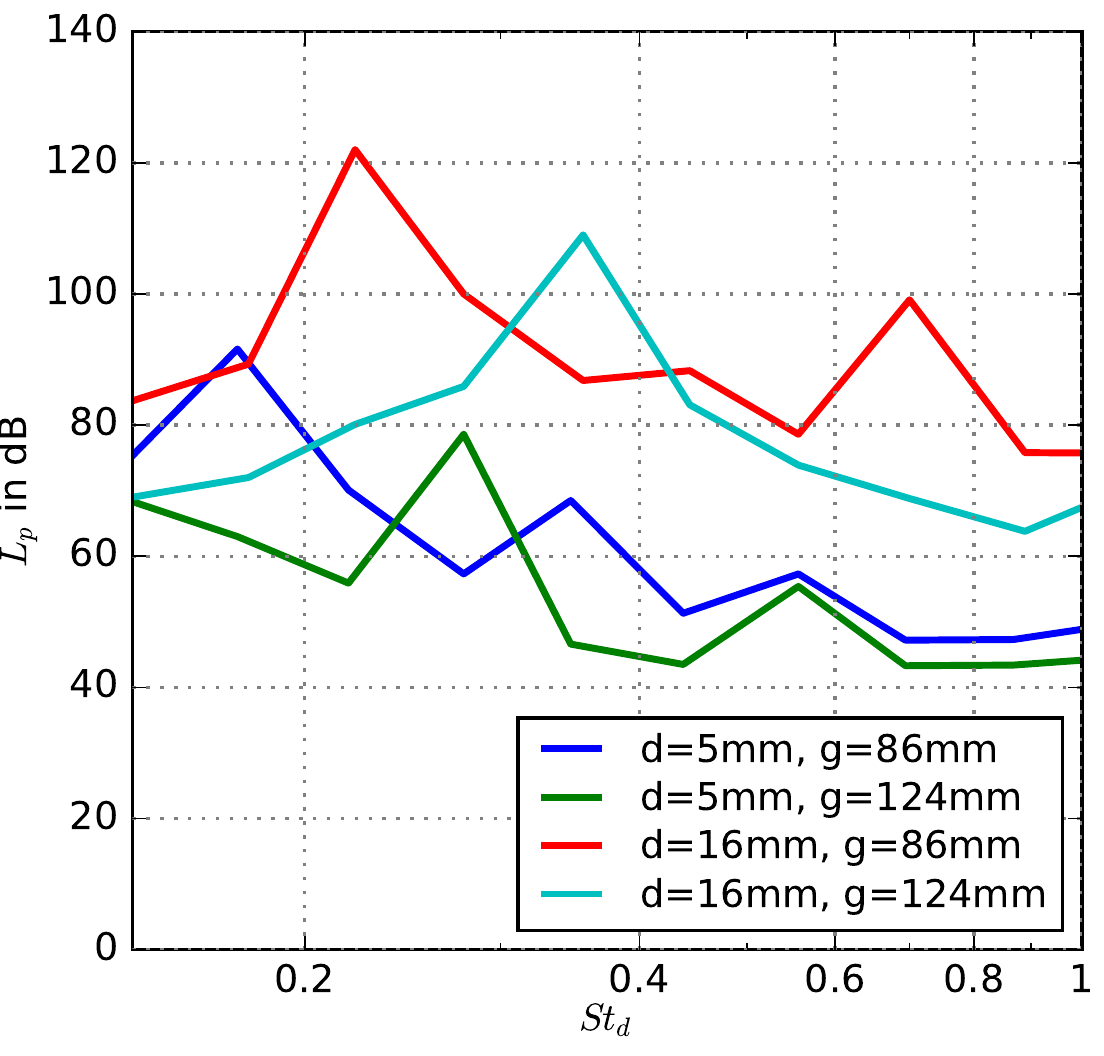}
            \caption[]%
            {}    
            \label{fig:fig5b}
        \end{subfigure}
        \caption{Numerically obtained 1/3 octave-band sound pressure level spectra for all cases from 2D-DDES. (a) $M=0.076$, (b) $M=0.210$.}
        \label{fig:fig5}
\end{figure*} 

\subsubsection{2D-DDES:}
\noindent Fig.~\ref{fig:fig5} displays the 1/3 octave-band spectra of the sound pressure levels for all the eight cases from Table~\ref{tab:table2} as a first summary of the 2D computations. It is clear that the 2D-DDES simulations are incapable of predicting the correct trend of the sound pressure level spectra for all cases. However, surprisingly, the simulations succeeded in finding out which case is yielding higher sound pressure levels and which one results in lower levels. It can be observed from Fig.~\ref{fig:fig5} that case-5 and 7 are emitting higher sound pressure levels compared to case-6 and 8. This is in good agreement with the experimental observations. However, it can be observed that it is not possible to extract the accurate flow physics and acoustics from the two-dimensional simulations, although it still seems possible to identify which rod-airfoil configuration is noisier and which is less.

\subsubsection{3D-DDES:}
\noindent
For the three-dimensional simulation, the sound pressure levels $L_p$ obtained from the numerical simulations have to be corrected due to the fact that the set-up spanwise length $b_{num}~=~3d$ in the numerical simulations is shorter than the experimental spanwise length $b_{exp}~=~7.5d$. These corrections have been developed by \cite{kato1991large,Seo2007,Moon2010} involving the pressure coherence function in the spanwise direction or the spanwise coherence length $b_c$. As the spanwise coherence length of the cylinder is calculated approximately 2.46$d$, a constant sound pressure level correction of +3.91 dB is added across the whole spectrum.

\begin{figure*}
\captionsetup[subfigure]{justification=centering}
        \centering
        \begin{subfigure}[b]{1\columnwidth}
            \centering
            \includegraphics[width=1\columnwidth]{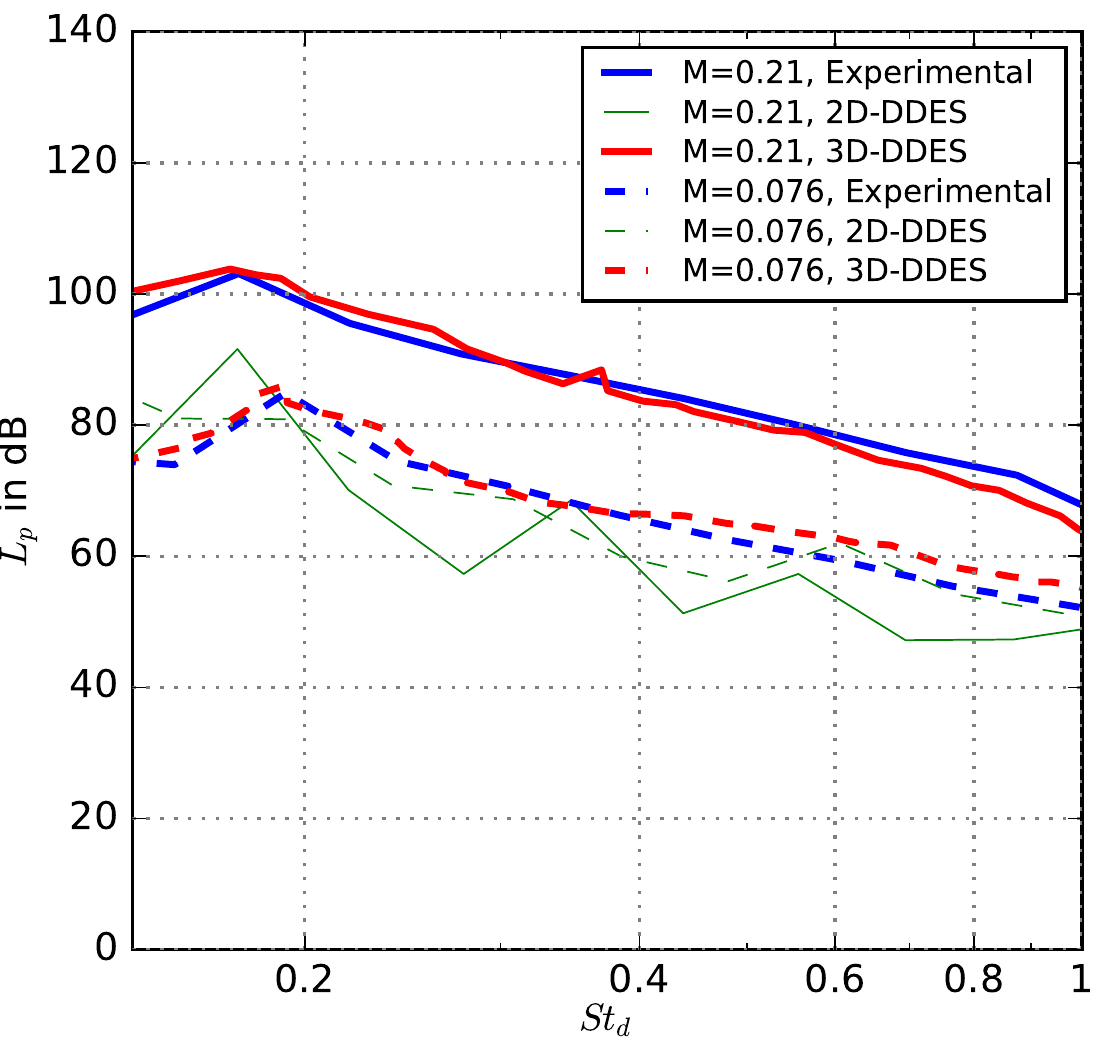}
            \caption[]%
            {} 
            \label{fig:fig10a}
        \end{subfigure}
        \hfill
        \begin{subfigure}[b]{1\columnwidth}   
            \centering
            \includegraphics[width=1\columnwidth]{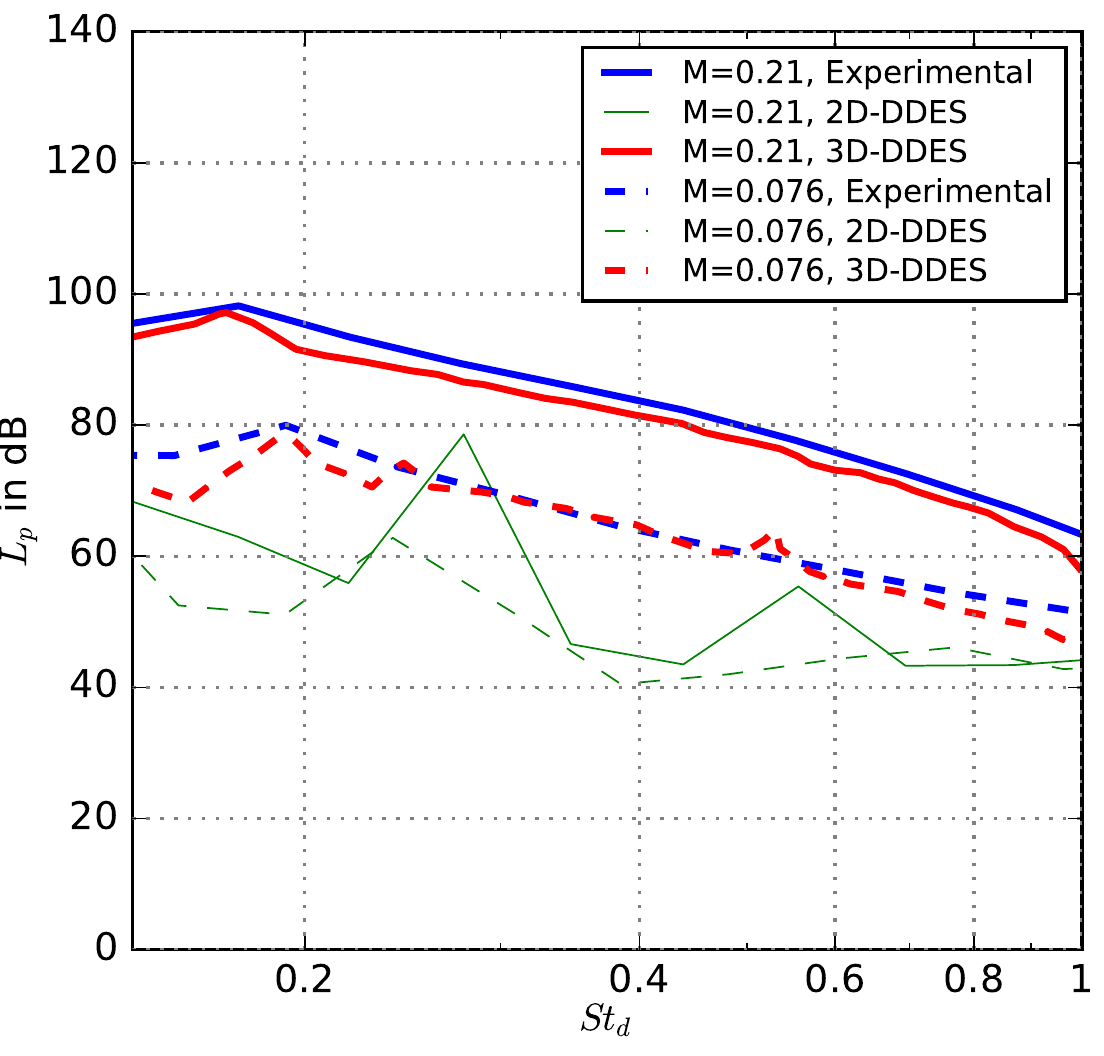}
            \caption[]%
            {} 
            \label{fig:fig10b}
        \end{subfigure}
        \begin{subfigure}[b]{1\columnwidth}
            \centering
            \includegraphics[width=1\columnwidth]{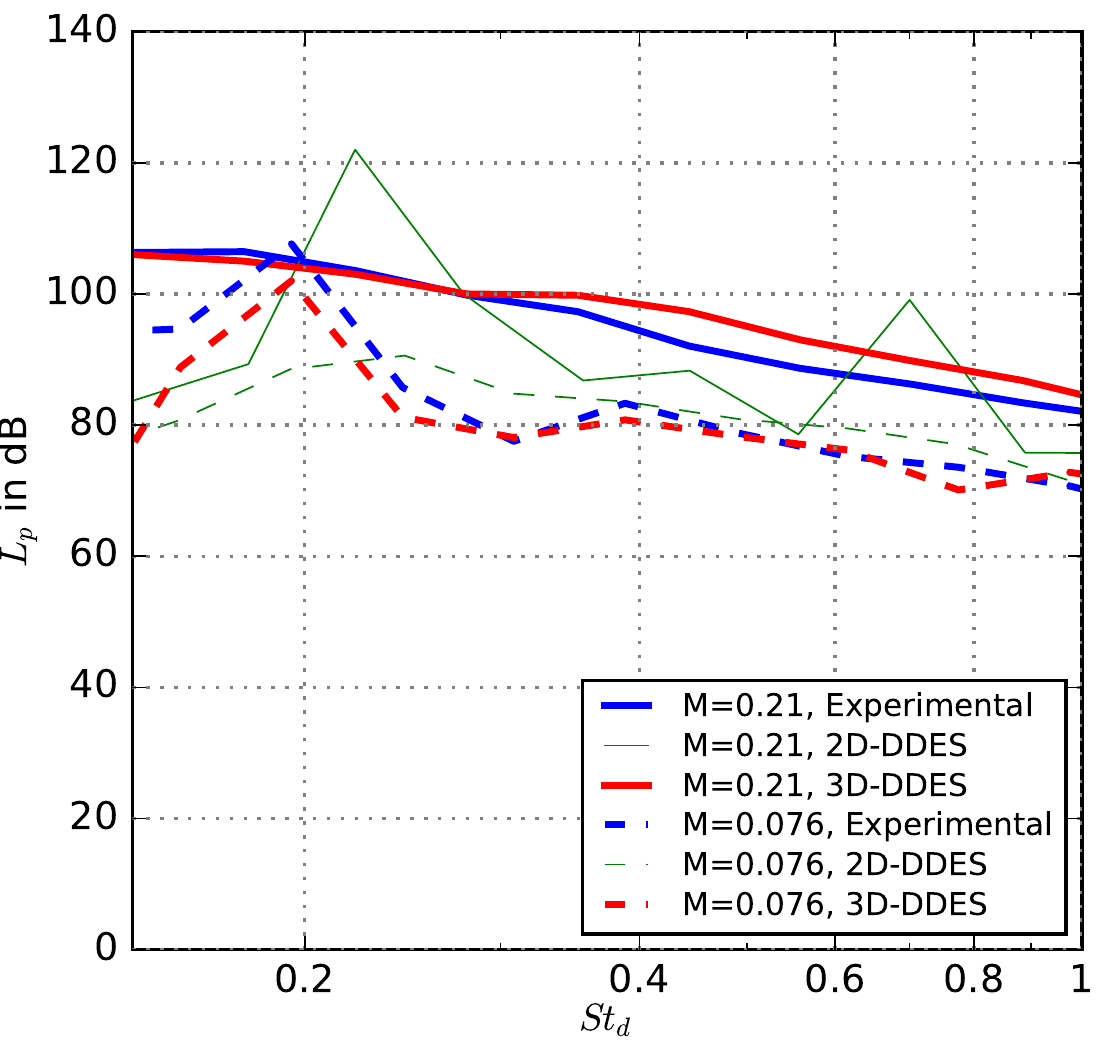}
            \caption[]%
            {} 
            \label{fig:fig10a}
        \end{subfigure}
        \hfill
        \begin{subfigure}[b]{1\columnwidth}   
            \centering
            \includegraphics[width=1\columnwidth]{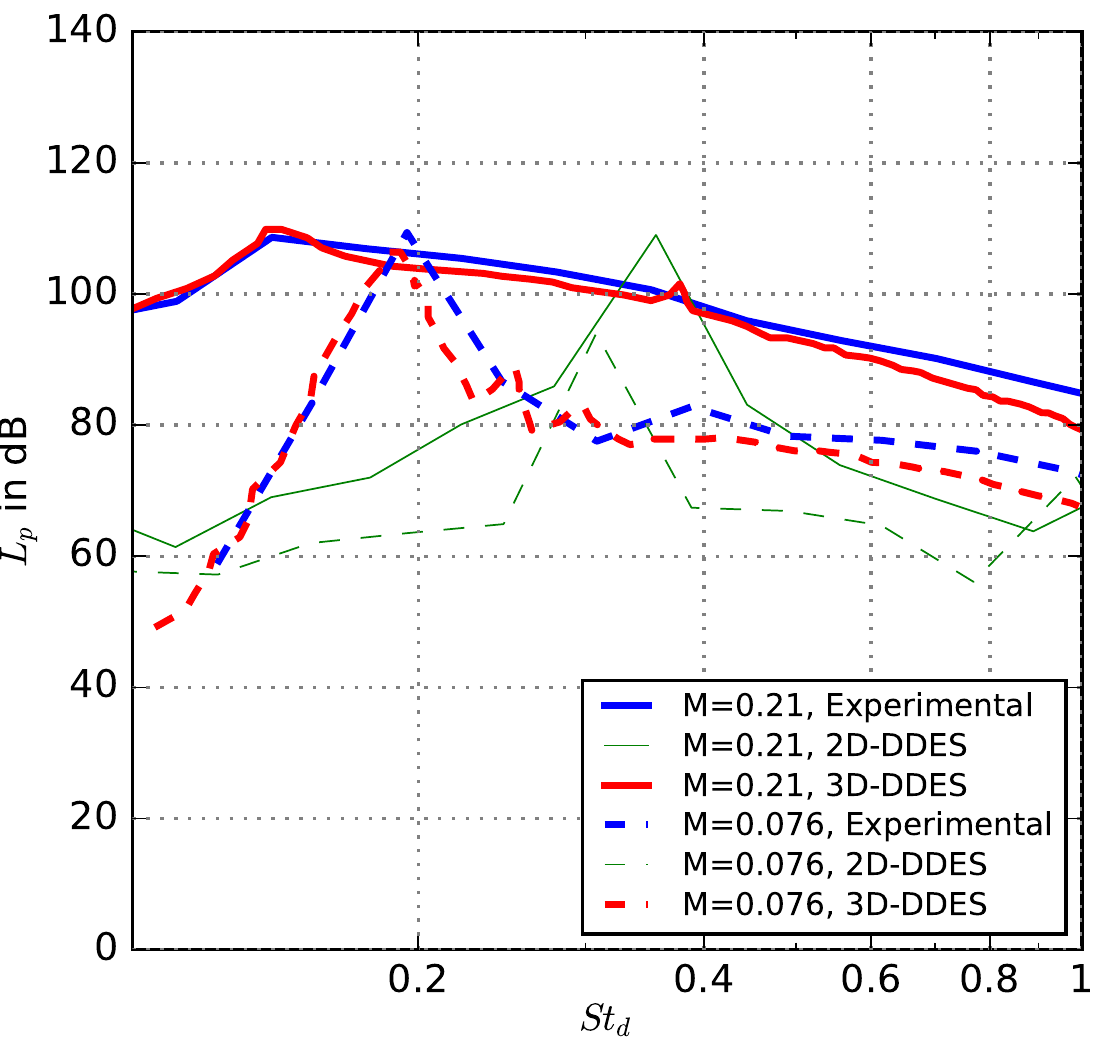}
            \caption[]%
            {} 
            \label{fig:fig10b}
        \end{subfigure}
        \caption{Comparison of Sound Pressure Level spectra in the far-field obtained obtained from 3D numerical results and experimental measurements. (a) Case-1 and Case-2, (b) Case-3 and Case-4, (c) Case-5 and Case-6, (d) Case-7 and Case-8.}
        \label{fig:fig10}
\end{figure*} 

\noindent The far-field noise at a position corresponding to the centre of the microphone array, located above the rod and the airfoil configuration (see Fig.~\ref{fig:schematic}), obtained from the 2D and 3D simulation is compared with that obtained from the experiments in Fig.~\ref{fig:fig10}. In general, the 3D numerical results are able to predict the frequency at the peak of the $L_p$ curve. From that, it can be said that the K\'arm\'an vortex shedding region and the stretch and the split of vortices near the leading edge of the airfoil are the primary acoustic source region. 
The disagreement of 2D acoustic results follows from the fact that 2D simulation of the hydrodynamic field predicts wrong acoustic sources. Typically, in experiment, a long-span test specimen is investigated and the spanwise decorrelation occurs. A good agreement between the experiments and 3D simulations can be observed in all the cases in Fig. \ref{fig:fig10} because only a 3D scale resolving simulation can capture such 3D noise source. However, it is possible to do 3D flow simulation and then 2D FWH but if the entire aeroacoustic simulation chain is in 2D, then it is necessary to apply empirical corrections before comparing the predicted noise levels to the measured levels. 

\section{Conclusion}
This paper investigates the effects of rod diameter and the streamwise gap between the rod and the leading edge of the airfoil on the noise generated by a rod–airfoil configuration. Experimental measurements and numerical simulations are conducted on eight cases of rod–airfoil configurations at two flow speeds, where the geometric parameters were varied by changing the rod diameter ($d$) and the streamwise gap ($g$) between the rod and the leading edge of the airfoil. The general conclusions are as follows:

\begin{enumerate}
    \item For massively separated turbulent flows, DDES is adopted and validated with the LES results. The far-field acoustic prediction using the FWH approach shows good agreement with the experimental data. 
    \item The sound pressure levels are highly dependent on the geometric parameters. The radiated noise increases with increasing rod diameter and/or shortening the streamwise gap for a thinner rod, while for the thicker rod diameter the larger streamwise gap generate slightly more noise.
    \item It was also found that 2D simulations can only predict a very basic spectral shape and trends of sound pressure levels observed in the experiments. The simulated sound pressure levels are usually lower than those measured in experiments.
\end{enumerate}

\subsection*{Acknowledgements}
The authors thank Erik W. Schneehagen and Ennes Sarradj from TU Berlin and Andreas Krebs from BTU Cottbus for their help with the three-dimensional simulations.

\appendix
\section{Appendix}
Kato’s formula, which assumes that the coherence function has the form of a boxcar function, is given by

\begin{equation}
    \text{If}~~ b_c(\omega) < b_{num},~~ (L_p)_{exp} = (L_p)_{num} + 10\log{\frac{b_{exp}}{b_{num}}}, 
\end{equation}

\begin{equation}
\begin{aligned}
    \text{If}~~ b_{num} < b_c(\omega) < b_{exp},{} &~~ (L_p)_{exp} = (L_p)_{num} \\ 
    & + 20\log\frac{b_{c}(\omega)}{b_{num}} + 10\log\frac{b_{exp}}{b_{c}(\omega)},
\end{aligned}
\end{equation}

\begin{equation}
    \text{If}~~ b_{exp} < b_c(\omega),~~ (L_p)_{exp} = (L_p)_{num} + 20log{\frac{b_{exp}}{b_{num}}}. 
\end{equation}

\noindent
According to \cite{kato1991large}, SPL correction for the long span can be made by adding $10 \log({b_{exp}}/{b_{num}})$ ,if a coherence length of the surface pressure fluctuations, $L_c$ is determined less than the simulated span. The equivalent coherence length, $L_c(\omega)$, can be obtained by calculating the coherence function
$\Gamma_{ij}(\omega)$ of the fluctuating wall pressure which is given by

\begin{equation}
    \Gamma_{ij}(\omega)=\frac{\overline{Re\left ( \hat{P'_i}\hat{P^*_j}  \right )}}{\sqrt{\left | \hat{P'_i} \right |^{2}}\sqrt{\left | \hat{P^*_j} \right |^{2}}},
    \label{eq:eq_}
\end{equation}

\noindent
where $\hat{P'_i}$ and $\hat{P^*_j}$ are pressure fluctuations (frequency domain) at different span positions on the cylinder surface. The cross-power spectrum $\hat{P'_i}\hat{P^*_j}$ can be evaluated with the surface pressure, using the Curle’s analogy solution

\begin{equation}
    \hat{P'}(\omega)\simeq \frac{1}{4\pi c_0}\int_{}^{}(\mathbf{r}\cdot\hat n)(-i\omega\hat P (\omega)){\rm{exp}}(i\omega r/c_0)dS.
    \label{eq:r2}
\end{equation}

\noindent
For a compact source or when the observer’s position is very far, it can be assumed that $r/c_0 \approx$ constant. Then, the cross-power spectrum $\hat{P'_i}\hat{P^*_j}$ is analytically written as

\begin{equation}
    \hat{P'_i}\hat{P^*_j}\simeq \left | \frac{i\omega\,{\rm exp}(i\omega r/c_0)}{4\pi c_0} \right |^{2}\int_{}^{}(\mathbf{r}\cdot\hat n)\hat P_i dS_i \cdot \int_{}^{}(\mathbf{r}\cdot\hat n)\hat P^*_j dS_j,
    \label{eq:r3}
\end{equation}

\noindent
and the acoustic spanwise coherence function can be expressed as

\begin{equation}
    {\gamma}'(\Delta z_{ij})\simeq \frac{\overline{Re\left ( \int_{}^{}(\mathbf{r}\cdot\hat n)\hat P_i dS_i \right )\cdot \int_{}^{}(\mathbf{r}\cdot\hat n)\hat P^*_j dS_j}}{\sqrt{\left | \int_{}^{}(\mathbf{r}\cdot\hat n)\hat P_i dS_i \right |^2}\sqrt{\left | \int_{}^{}(\mathbf{r}\cdot\hat n)\hat P^*_j dS_j \right |^{2}}}, 
\end{equation}

\noindent
where $\hat P_i$ is the surface pressure at each subsection and $\int_{}^{}dS_i$ is the
surface integral over each sub-sectional area. Eq.~\eqref{eq:r3} is the relation
between the acoustic spanwise coherence function, ${\gamma}'(\Delta z_{ij})$ and the
spanwise coherence function of the ‘integrated’ surface pressure. Using above equation, the coherence length $b_c$ is defined as

\begin{equation}
    b_c = \int_{z_1}^{+\infty}{\gamma}'(\Delta z_{12})dz_2.
\end{equation}

\bibliographystyle{model1-num-names}

\bibliography{cas-refs}

\begin{thebibliography}{58}
\expandafter\ifx\csname natexlab\endcsname\relax\def\natexlab#1{#1}\fi
\providecommand{\url}[1]{\texttt{#1}}
\providecommand{\href}[2]{#2}
\providecommand{\path}[1]{#1}
\providecommand{\DOIprefix}{doi:}
\providecommand{\ArXivprefix}{arXiv:}
\providecommand{\URLprefix}{URL: }
\providecommand{\Pubmedprefix}{pmid:}
\providecommand{\doi}[1]{\href{http://dx.doi.org/#1}{\path{#1}}}
\providecommand{\Pubmed}[1]{\href{pmid:#1}{\path{#1}}}
\providecommand{\bibinfo}[2]{#2}
\ifx\xfnm\relax \def\xfnm[#1]{\unskip,\space#1}\fi
\bibitem[{Blake(1986)}]{Blake1986}
\bibinfo{author}{W.~K. Blake}, \bibinfo{title}{Mechanics of Flow-Induced Sound
  and Vibration, Volume II: Complex Flow-Structure Interactions},
  \bibinfo{publisher}{Academic Press, Inc.}, \bibinfo{year}{1986}.
\bibitem[{Schulten(1997)}]{Schulten1997}
\bibinfo{author}{J.~B. Schulten},
\newblock \bibinfo{title}{{Vane sweep effects on rotor/stator interaction
  noise}},
\newblock \bibinfo{journal}{AIAA Journal} \bibinfo{volume}{35}
  (\bibinfo{year}{1997}) \bibinfo{pages}{945--951}.
\bibitem[{Cooper and Peake(2005)}]{Cooper2005}
\bibinfo{author}{A.~J. Cooper}, \bibinfo{author}{N.~Peake},
\newblock \bibinfo{title}{{Upstream-radiated rotor-stator interaction noise in
  mean swirling flow}},
\newblock \bibinfo{journal}{Journal of Fluid Mechanics} \bibinfo{volume}{523}
  (\bibinfo{year}{2005}) \bibinfo{pages}{219--250}.
\bibitem[{Cooper and Peake(2006)}]{Cooper2006}
\bibinfo{author}{A.~J. Cooper}, \bibinfo{author}{N.~Peake},
\newblock \bibinfo{title}{{Rotor-stator interaction noise in swirling flow:
  Stator sweep and lean effects}},
\newblock \bibinfo{journal}{AIAA Journal} \bibinfo{volume}{44}
  (\bibinfo{year}{2006}) \bibinfo{pages}{981--991}.
\bibitem[{Jacob et~al.(2005)Jacob, Boudet, Casalino, and Michard}]{Jacob2005}
\bibinfo{author}{M.~C. Jacob}, \bibinfo{author}{J.~Boudet},
  \bibinfo{author}{D.~Casalino}, \bibinfo{author}{M.~Michard},
\newblock \bibinfo{title}{{A rod-airfoil experiment as a benchmark for
  broadband noise modeling}},
\newblock \bibinfo{journal}{Theoretical and Computational Fluid Dynamics}
  \bibinfo{volume}{19} (\bibinfo{year}{2005}) \bibinfo{pages}{171--196}.
\bibitem[{Casalino et~al.(2003)Casalino, Jacob, and Roger}]{Casalino2003}
\bibinfo{author}{D.~Casalino}, \bibinfo{author}{M.~Jacob},
  \bibinfo{author}{M.~Roger},
\newblock \bibinfo{title}{{Prediction of Rod-Airfoil Interaction Noise Using
  the Ffowcs-Williams-Hawkings Analogy}},
\newblock \bibinfo{journal}{AIAA Journal} \bibinfo{volume}{41}
  (\bibinfo{year}{2003}) \bibinfo{pages}{182--191}.
\bibitem[{Ffowcs~Williams and Hawkings(1969)}]{Ffowcs1969}
\bibinfo{author}{J.~E. Ffowcs~Williams}, \bibinfo{author}{D.~L. Hawkings},
\newblock \bibinfo{title}{Sound generation by turbulence and surfaces in
  arbitrary motion},
\newblock \bibinfo{journal}{Philosophical Transactions for the Royal Society of
  London. Series A, Mathematical and Physical Sciences}  (\bibinfo{year}{1969})
  \bibinfo{pages}{321--342}.
\bibitem[{Boudet et~al.(2005)Boudet, Grosjean, and Jacob}]{Boudet2005}
\bibinfo{author}{J.~Boudet}, \bibinfo{author}{N.~Grosjean},
  \bibinfo{author}{M.~C. Jacob},
\newblock \bibinfo{title}{{Wake-Airfoil Interaction as Broadband Noise Source:
  A Large-Eddy Simulation Study}},
\newblock \bibinfo{journal}{International Journal of Aeroacoustics}
  \bibinfo{volume}{4} (\bibinfo{year}{2005}) \bibinfo{pages}{93--115}.
\bibitem[{Berland et~al.(2010)Berland, Lafon, Crouzet, Daude, and
  Bailly}]{Berland2010}
\bibinfo{author}{J.~Berland}, \bibinfo{author}{P.~Lafon},
  \bibinfo{author}{F.~Crouzet}, \bibinfo{author}{F.~Daude},
  \bibinfo{author}{C.~Bailly},
\newblock \bibinfo{title}{{Numerical Insight into Sound Sources of a
  Rod-Airfoil Flow Configuration Using Direct Noise Calculation}},
\newblock in: \bibinfo{booktitle}{16th AIAA/CEAS Aeroacoustics Conference},
  \bibinfo{year}{2010}.
\bibitem[{Giret et~al.(2012)Giret, Sengissen, Moreau, Sanjos{\'{e}}, and
  Jouhaud}]{Giret2012}
\bibinfo{author}{J.-C. Giret}, \bibinfo{author}{A.~Sengissen},
  \bibinfo{author}{S.~Moreau}, \bibinfo{author}{M.~Sanjos{\'{e}}},
  \bibinfo{author}{J.-c. Jouhaud}, \bibinfo{title}{{Prediction of the sound
  generated by a rod-airfoil configuration using a compressible unstructured
  LES solver and a FW-H analogy}}, \bibinfo{year}{2012}.
  \DOIprefix\doi{10.2514/6.2012-2058}.
\bibitem[{Agrawal and Sharma(2016)}]{Agrawal2016}
\bibinfo{author}{B.~R. Agrawal}, \bibinfo{author}{A.~Sharma},
\newblock \bibinfo{title}{Numerical analysis of aerodynamic noise mitigation
  via leading edge serrations for a rod--airfoil configuration},
\newblock \bibinfo{journal}{International Journal of Aeroacoustics}
  \bibinfo{volume}{15} (\bibinfo{year}{2016}) \bibinfo{pages}{734--756}.
\bibitem[{Jiang et~al.(2015)Jiang, Mao, Deng, and Liu}]{Jiang2015}
\bibinfo{author}{Y.~Jiang}, \bibinfo{author}{M.~L. Mao}, \bibinfo{author}{X.~G.
  Deng}, \bibinfo{author}{H.~Y. Liu},
\newblock \bibinfo{title}{{Numerical investigation on body-wake flow
  interaction over rod-airfoil configuration}},
\newblock \bibinfo{journal}{Journal of Fluid Mechanics} \bibinfo{volume}{779}
  (\bibinfo{year}{2015}) \bibinfo{pages}{1--35}.
\bibitem[{Zhiyin(2015)}]{Zhiyin2015}
\bibinfo{author}{Y.~Zhiyin}, \bibinfo{title}{{Large-eddy simulation: Past,
  present and the future}}, \bibinfo{year}{2015}.
  \DOIprefix\doi{10.1016/j.cja.2014.12.007}.
\bibitem[{Geyer et~al.(2018)Geyer, Sharma, and Sarradj}]{Geyer2018}
\bibinfo{author}{T.~F. Geyer}, \bibinfo{author}{S.~Sharma},
  \bibinfo{author}{E.~Sarradj},
\newblock \bibinfo{title}{{Detached Eddy Simulation of the Flow Noise
  Generation of Cylinders with Porous Cover}},
\newblock in: \bibinfo{booktitle}{2018 AIAA/CEAS Aeroacoustics Conference, AIAA
  paper 2018-3472}, \bibinfo{year}{2018}.
\bibitem[{Greschner et~al.(2004)Greschner, Thiele, Casalino, and
  Jacob}]{Greschner2004}
\bibinfo{author}{B.~Greschner}, \bibinfo{author}{F.~Thiele},
  \bibinfo{author}{D.~Casalino}, \bibinfo{author}{M.~Jacob},
\newblock \bibinfo{title}{{Influence of Turbulence Modeling on the Broadband
  Noise Simulation for Complex Flows}},
\newblock in: \bibinfo{booktitle}{10th AIAA/CEAS Aeroacoustics Conference},
  \bibinfo{year}{2004}.
\bibitem[{Zhou et~al.(2017)Zhou, Albring, Gauger, Ilario, Economon, and
  Alonso}]{Zhou2017}
\bibinfo{author}{B.~Zhou}, \bibinfo{author}{T.~A. Albring},
  \bibinfo{author}{N.~R. Gauger}, \bibinfo{author}{C.~Ilario},
  \bibinfo{author}{T.~D. Economon}, \bibinfo{author}{J.~J. Alonso},
\newblock \bibinfo{title}{Reduction of airframe noise components using a
  discrete adjoint approach},
\newblock in: \bibinfo{booktitle}{18th AIAA/ISSMO Multidisciplinary Analysis
  and Optimization Conference, AIAA paper 2017-3658}, \bibinfo{year}{2017}.
\bibitem[{Galdeano et~al.(2010)Galdeano, Barr{\'{e}}, and
  R{\'{e}}au}]{Galdeano2010}
\bibinfo{author}{S.~Galdeano}, \bibinfo{author}{S.~Barr{\'{e}}},
  \bibinfo{author}{N.~R{\'{e}}au},
\newblock \bibinfo{title}{{Noise radiated by a rod-airfoil configuration using
  DES and the Ffowcs-Williams {\&} Hawkings' analogy}},
\newblock in: \bibinfo{booktitle}{16th AIAA/CEAS Aeroacoustics Conference, AIAA
  paper 2010-3702}, \bibinfo{year}{2010}.
\bibitem[{Greschner et~al.(2008)Greschner, Thiele, Jacob, and
  Casalino}]{Greschner2008}
\bibinfo{author}{B.~Greschner}, \bibinfo{author}{F.~Thiele},
  \bibinfo{author}{M.~C. Jacob}, \bibinfo{author}{D.~Casalino},
\newblock \bibinfo{title}{{Prediction of sound generated by a rod–airfoil
  configuration using EASM DES and the generalised Lighthill/FW-H analogy}},
\newblock \bibinfo{journal}{Computers {\&} Fluids} \bibinfo{volume}{37}
  (\bibinfo{year}{2008}) \bibinfo{pages}{402--413}.
\bibitem[{Caraeni et~al.(2007)Caraeni, Dai, and Caraeni}]{Caraeni2007}
\bibinfo{author}{M.~Caraeni}, \bibinfo{author}{Y.~Dai},
  \bibinfo{author}{D.~Caraeni},
\newblock \bibinfo{title}{{Acoustic Investigation of Rod Airfoil Configuration
  with DES and FWH}},
\newblock in: \bibinfo{booktitle}{37th AIAA Fluid Dynamics Conference and
  Exhibit, AIAA paper 2007-4106}, \bibinfo{year}{2007}.
\bibitem[{Gerolymos and Vallet(2007)}]{Gerolymos2007}
\bibinfo{author}{G.~A. Gerolymos}, \bibinfo{author}{I.~Vallet},
\newblock \bibinfo{title}{{Influence of Temporal Integration and Spatial
  Discretization on Hybrid RSM-VLES Computations}},
\newblock in: \bibinfo{booktitle}{18th AIAA Computational Fluid Dynamics
  Conference, AIAA paper 2007-4094}, \bibinfo{year}{2007}.
\bibitem[{Giesler and Sarradj(2009)}]{Giesler2009}
\bibinfo{author}{J.~Giesler}, \bibinfo{author}{E.~Sarradj},
\newblock \bibinfo{title}{{Measurement of broadband noise generation on
  rod-airfoil-configurations}},
\newblock in: \bibinfo{booktitle}{15th AIAA/CEAS Aeroacoustics Conference (30th
  AIAA Aeroacoustics Conference), AIAA paper 2009-3308}, \bibinfo{year}{2009}.
\bibitem[{Sarradj et~al.(2009)Sarradj, Fritzsche, Geyer, and
  Giesler}]{Sarradj2009}
\bibinfo{author}{E.~Sarradj}, \bibinfo{author}{C.~Fritzsche},
  \bibinfo{author}{T.~F. Geyer}, \bibinfo{author}{J.~Giesler},
\newblock \bibinfo{title}{{Acoustic and aerodynamic design and characterization
  of a small-scale aeroacoustic wind tunnel}},
\newblock \bibinfo{journal}{Applied Acoustics} \bibinfo{volume}{70}
  (\bibinfo{year}{2009}) \bibinfo{pages}{1073--1080}.
\bibitem[{Giesler(2011)}]{Giesler2011}
\bibinfo{author}{J.~Giesler}, \bibinfo{title}{Schallentstehung durch turbulente
  Zustr{\"o}mung an aerodynamischen Profilen}, \bibinfo{type}{Doctoral thesis},
  Brandenburg University of Technology, Cottbus, \bibinfo{year}{2011}.
\bibitem[{Sarradj and Herold(2017)}]{sarradj2017python}
\bibinfo{author}{E.~Sarradj}, \bibinfo{author}{G.~Herold},
\newblock \bibinfo{title}{A python framework for microphone array data
  processing},
\newblock \bibinfo{journal}{Applied Acoustics} \bibinfo{volume}{116}
  (\bibinfo{year}{2017}) \bibinfo{pages}{50--58}.
\bibitem[{Welch(1967)}]{welch1967use}
\bibinfo{author}{P.~Welch},
\newblock \bibinfo{title}{The use of fast fourier transform for the estimation
  of power spectra: a method based on time averaging over short, modified
  periodograms},
\newblock \bibinfo{journal}{IEEE Transactions on audio and electroacoustics}
  \bibinfo{volume}{15} (\bibinfo{year}{1967}) \bibinfo{pages}{70--73}.
\bibitem[{Brooks and Humphreys(2006)}]{Brooks2006}
\bibinfo{author}{T.~F. Brooks}, \bibinfo{author}{W.~M. Humphreys},
\newblock \bibinfo{title}{{A deconvolution approach for the mapping of acoustic
  sources (DAMAS) determined from phased microphone arrays}},
\newblock \bibinfo{journal}{Journal of Sound and Vibration}
  \bibinfo{volume}{294} (\bibinfo{year}{2006}) \bibinfo{pages}{856--879}.
\bibitem[{Herold and Sarradj(2017)}]{Herold2017}
\bibinfo{author}{G.~Herold}, \bibinfo{author}{E.~Sarradj},
\newblock \bibinfo{title}{Performance analysis of microphone array methods},
\newblock \bibinfo{journal}{Journal of Sound and Vibration}
  \bibinfo{volume}{401} (\bibinfo{year}{2017}) \bibinfo{pages}{152--168}.
\bibitem[{Merino-Martinez et~al.(2019)Merino-Martinez, Sijtsma, Snellen,
  Ahlefeld, Antoni, Bahr, Blacodon, Ernst, Finez, Funke, Geyer, Haxter, Herold,
  Huang, Humphreys, Leclère, Malgoezar, Michel, Padois, Pereira, Picard,
  Sarradj, Siller, Simons, and Spehr}]{Merino2019}
\bibinfo{author}{R.~Merino-Martinez}, \bibinfo{author}{P.~Sijtsma},
  \bibinfo{author}{M.~Snellen}, \bibinfo{author}{T.~Ahlefeld},
  \bibinfo{author}{J.~Antoni}, \bibinfo{author}{C.~J. Bahr},
  \bibinfo{author}{D.~Blacodon}, \bibinfo{author}{D.~Ernst},
  \bibinfo{author}{A.~Finez}, \bibinfo{author}{S.~Funke},
  \bibinfo{author}{T.~F. Geyer}, \bibinfo{author}{S.~Haxter},
  \bibinfo{author}{G.~Herold}, \bibinfo{author}{X.~Huang},
  \bibinfo{author}{W.~M. Humphreys}, \bibinfo{author}{Q.~Leclère},
  \bibinfo{author}{A.~Malgoezar}, \bibinfo{author}{U.~Michel},
  \bibinfo{author}{T.~Padois}, \bibinfo{author}{A.~Pereira},
  \bibinfo{author}{C.~Picard}, \bibinfo{author}{E.~Sarradj},
  \bibinfo{author}{H.~Siller}, \bibinfo{author}{D.~G. Simons},
  \bibinfo{author}{C.~Spehr},
\newblock \bibinfo{title}{A review of acoustic imaging methods using phased
  microphone arrays},
\newblock \bibinfo{journal}{CEAS Aeronautical Journal}  (\bibinfo{year}{2019})
  \bibinfo{pages}{1--34}.
\bibitem[{Sarradj(2012)}]{sarradj2012three}
\bibinfo{author}{E.~Sarradj},
\newblock \bibinfo{title}{Three-dimensional acoustic source mapping with
  different beamforming steering vector formulations},
\newblock \bibinfo{journal}{Advances in Acoustics and Vibration}
  \bibinfo{volume}{2012} (\bibinfo{year}{2012}).
\bibitem[{Schlichting and Gersten(1997)}]{Schlichting1997}
\bibinfo{author}{H.~Schlichting}, \bibinfo{author}{K.~Gersten},
  \bibinfo{title}{Boundary-layer theory}, \bibinfo{edition}{9th edition} ed.,
  \bibinfo{publisher}{Springer Science+Business Media, Berlin},
  \bibinfo{year}{1997}.
\bibitem[{Spalart et~al.(1997)Spalart, Jou, Strelets, and
  Allmaras}]{Spalart1997}
\bibinfo{author}{P.~R. Spalart}, \bibinfo{author}{W.~H. Jou},
  \bibinfo{author}{M.~K. Strelets}, \bibinfo{author}{S.~R. Allmaras},
\newblock \bibinfo{title}{{Comments on the feasibility of LES for wings and on
  a hybrid RANS/LES approach}},
\newblock \bibinfo{journal}{Advances in DNS/LES} \bibinfo{volume}{1}
  (\bibinfo{year}{1997}) \bibinfo{pages}{4--8}.
\bibitem[{Spalart(2009)}]{Spalart2009}
\bibinfo{author}{P.~R. Spalart},
\newblock \bibinfo{title}{Detached-eddy simulation},
\newblock \bibinfo{journal}{Annual Review of Fluid Mechanics}
  \bibinfo{volume}{41} (\bibinfo{year}{2009}) \bibinfo{pages}{181--202}.
\bibitem[{Spalart et~al.(2006)Spalart, Deck, Shur, Squires, Strelets, and
  Travin}]{Spalart2006}
\bibinfo{author}{P.~R. Spalart}, \bibinfo{author}{S.~Deck},
  \bibinfo{author}{M.~L. Shur}, \bibinfo{author}{K.~D. Squires},
  \bibinfo{author}{M.~K. Strelets}, \bibinfo{author}{A.~Travin},
\newblock \bibinfo{title}{{A new version of detached-eddy simulation, resistant
  to ambiguous grid densities}},
\newblock \bibinfo{journal}{Theoretical and Computational Fluid Dynamics}
  \bibinfo{volume}{20} (\bibinfo{year}{2006}) \bibinfo{pages}{181--195}.
\bibitem[{Spalart and Allmaras(1992)}]{SAmodel}
\bibinfo{author}{P.~R. Spalart}, \bibinfo{author}{S.~R. Allmaras},
\newblock \bibinfo{title}{A one-equation turbulence model for aerodynamic
  flows}  (\bibinfo{year}{1992}).
\bibitem[{Smagorinsky(1963)}]{smagorinsky}
\bibinfo{author}{J.~Smagorinsky},
\newblock \bibinfo{title}{{General circulation experiments with the primitive
  equations: i. The basic experiment}},
\newblock \bibinfo{journal}{Monthly Weather Review} \bibinfo{volume}{91}
  (\bibinfo{year}{1963}) \bibinfo{pages}{99--164}.
\bibitem[{Economon et~al.(2015)Economon, Palacios, Copeland, Lukaczyk, and
  Alonso}]{Economon2015}
\bibinfo{author}{T.~D. Economon}, \bibinfo{author}{F.~Palacios},
  \bibinfo{author}{S.~R. Copeland}, \bibinfo{author}{T.~W. Lukaczyk},
  \bibinfo{author}{J.~J. Alonso},
\newblock \bibinfo{title}{Su2: An open-source suite for multiphysics simulation
  and design},
\newblock \bibinfo{journal}{Aiaa Journal} \bibinfo{volume}{54}
  (\bibinfo{year}{2015}) \bibinfo{pages}{828--846}.
\bibitem[{Molina et~al.(2017)Molina, Spode, Annes~da Silva, Manosalvas-Kjono,
  Nimmagadda, Economon, Alonso, and Righi}]{Molina2017}
\bibinfo{author}{E.~Molina}, \bibinfo{author}{C.~Spode}, \bibinfo{author}{R.~G.
  Annes~da Silva}, \bibinfo{author}{D.~E. Manosalvas-Kjono},
  \bibinfo{author}{S.~Nimmagadda}, \bibinfo{author}{T.~D. Economon},
  \bibinfo{author}{J.~J. Alonso}, \bibinfo{author}{M.~Righi},
\newblock \bibinfo{title}{Hybrid rans/les calculations in su2},
\newblock in: \bibinfo{booktitle}{23rd AIAA Computational Fluid Dynamics
  Conference, AIAA paper 2017-4284}, \bibinfo{year}{2017}.
\bibitem[{Jameson and Shankaran(2009)}]{Jameson}
\bibinfo{author}{A.~Jameson}, \bibinfo{author}{S.~Shankaran},
\newblock \bibinfo{title}{An assessment of dual-time stepping, time spectral
  and artificial compressibility based numerical algorithms for unsteady flow
  with applications to flapping wings}  (\bibinfo{year}{2009}).
\bibitem[{Farassat(2007)}]{Farassat2007}
\bibinfo{author}{F.~Farassat},
\newblock \bibinfo{title}{{Derivation of Formulations 1 and 1A of Farassat}},
\newblock \bibinfo{journal}{Nasa/TM-2007-214853} \bibinfo{volume}{214853}
  (\bibinfo{year}{2007}) \bibinfo{pages}{1--25}.
\bibitem[{Brentner and Farassat(2003)}]{BRENTNER200383}
\bibinfo{author}{K.~S. Brentner}, \bibinfo{author}{F.~Farassat},
\newblock \bibinfo{title}{Modeling aerodynamically generated sound of
  helicopter rotors},
\newblock \bibinfo{journal}{Progress in Aerospace Sciences}
  \bibinfo{volume}{39} (\bibinfo{year}{2003}) \bibinfo{pages}{83 -- 120}.
\bibitem[{Sharma et~al.(2020)Sharma, Sarradj, and Schmidt}]{Sharma2020a}
\bibinfo{author}{S.~Sharma}, \bibinfo{author}{E.~Sarradj},
  \bibinfo{author}{H.~Schmidt},
\newblock \bibinfo{title}{{Stochastic modelling of leading-edge noise in
  time-domain using vortex particles}},
\newblock \bibinfo{journal}{Journal of Sound and Vibration}
  (\bibinfo{year}{2020}) \bibinfo{pages}{1--175}.
\bibitem[{Sharma(2020)}]{Sharma2020}
\bibinfo{author}{S.~Sharma}, \bibinfo{title}{Stochastic modelling of
  leading-edge noise in time-domain using vortex particles},
  \bibinfo{type}{Doctoral thesis}, BTU Cottbus - Senftenberg,
  \bibinfo{year}{2020}. \DOIprefix\doi{10.26127/BTUOpen-5085}.
\bibitem[{Ewert et~al.(2009)Ewert, Appel, Dierke, and Herr}]{Ewert2009}
\bibinfo{author}{R.~Ewert}, \bibinfo{author}{C.~Appel},
  \bibinfo{author}{J.~Dierke}, \bibinfo{author}{M.~Herr},
\newblock \bibinfo{title}{{RANS/CAA Based Prediction of NACA 0012 Broadband
  Trailing Edge Noise and Experimental Validation}},
\newblock in: \bibinfo{booktitle}{15th AIAA/CEAS Aeroacoustics Conference (30th
  AIAA Aeroacoustics Conference)}, \bibinfo{publisher}{American Institute of
  Aeronautics and Astronautics}, \bibinfo{address}{Reston, Virigina},
  \bibinfo{year}{2009}. \DOIprefix\doi{10.2514/6.2009-3269}.
\bibitem[{Amiet(1976)}]{Amiet1976}
\bibinfo{author}{R.~K. Amiet},
\newblock \bibinfo{title}{{Noise due to turbulent flow past a trailing edge}},
\newblock \bibinfo{journal}{Journal of Sound and Vibration}
  \bibinfo{volume}{47} (\bibinfo{year}{1976}) \bibinfo{pages}{387--393}.
\bibitem[{Zdravkovich(1997)}]{Zdravkovich1997}
\bibinfo{author}{M.~M. Zdravkovich}, \bibinfo{title}{{Flow around circular
  cylinders: A comprehensive guide through flow phenomena, experiments,
  applications, mathematical models, and computer simulations}},
  \bibinfo{publisher}{Oxford University Press}, \bibinfo{year}{1997}.
\bibitem[{Sharma et~al.(2019)Sharma, Geyer, Sarradj, and Schmidt}]{Sharma2019}
\bibinfo{author}{S.~Sharma}, \bibinfo{author}{T.~F. Geyer},
  \bibinfo{author}{E.~Sarradj}, \bibinfo{author}{H.~Schmidt},
\newblock \bibinfo{title}{Numerical investigation of noise generation by
  rod-airfoil configuration using des (su2) and the fw-h analogy},
\newblock in: \bibinfo{booktitle}{25th AIAA/CEAS Aeroacoustics Conference},
  \bibinfo{year}{2019}. \DOIprefix\doi{10.2514/6.2019-2400}.
\bibitem[{Mockett(2009)}]{Mockett2009}
\bibinfo{author}{C.~Mockett}, \bibinfo{title}{{A comprehensive study of
  detached-eddy simulation}}, Ph.D. thesis, Technische Universit{\"{a}}t
  Berlin, \bibinfo{year}{2009}. \URLprefix
  \url{https://depositonce.tu-berlin.de/handle/11303/2602}.
\bibitem[{Gershfeld(2004)}]{Gershfeld2004b}
\bibinfo{author}{J.~Gershfeld},
\newblock \bibinfo{title}{{Leading edge noise from thick foils in turbulent
  flows}},
\newblock \bibinfo{journal}{Journal of the Acoustical Society of America}
  \bibinfo{volume}{116} (\bibinfo{year}{2004}) \bibinfo{pages}{1416--1426}.
\bibitem[{Li et~al.(2014)Li, nian Wang, wu~Chen, and chu Li}]{Li2014}
\bibinfo{author}{Y.~Li}, \bibinfo{author}{X.~nian Wang},
  \bibinfo{author}{Z.~wu~Chen}, \bibinfo{author}{Z.~chu Li},
\newblock \bibinfo{title}{{Experimental study of vortex-structure interaction
  noise radiated from rod-airfoil configurations}},
\newblock \bibinfo{journal}{Journal of Fluids and Structures}
  \bibinfo{volume}{51} (\bibinfo{year}{2014}) \bibinfo{pages}{313--325}.
\bibitem[{Amiet(1975)}]{Amiet1975}
\bibinfo{author}{R.~K. Amiet},
\newblock \bibinfo{title}{Acoustic radiation from an airfoil in a turbulent
  stream},
\newblock \bibinfo{journal}{Journal of Sound and vibration}
  \bibinfo{volume}{41} (\bibinfo{year}{1975}) \bibinfo{pages}{407--420}.
\bibitem[{Geyer(2018)}]{Geyer2018B}
\bibinfo{author}{T.~F. Geyer}, \bibinfo{title}{{Motor- und Aggregate-Akustik,
  10. Magdeburger Symposium}},
  \bibinfo{publisher}{Otto-von-Guericke-Universität Magdeburg},
  \bibinfo{year}{2018}, pp. \bibinfo{pages}{185 -- 206}.
  \DOIprefix\doi{10.24352/UB.OVGU-2018-115}.
\bibitem[{Roach(1987)}]{Roach1987}
\bibinfo{author}{P.~E. Roach},
\newblock \bibinfo{title}{The generation of nearly isotropic turbulence by
  means of grids},
\newblock \bibinfo{journal}{International Journal of Heat and Fluid Flow}
  \bibinfo{volume}{8} (\bibinfo{year}{1987}) \bibinfo{pages}{82--92}.
\bibitem[{Lockard et~al.(2007)Lockard, Khorrami, Choudhari, Hutcheson, Brooks,
  and Stead}]{lockard2007tandem}
\bibinfo{author}{D.~Lockard}, \bibinfo{author}{M.~Khorrami},
  \bibinfo{author}{M.~Choudhari}, \bibinfo{author}{F.~Hutcheson},
  \bibinfo{author}{T.~Brooks}, \bibinfo{author}{D.~Stead},
\newblock \bibinfo{title}{Tandem cylinder noise predictions},
\newblock in: \bibinfo{booktitle}{13th AIAA/CEAS Aeroacoustics Conference (28th
  AIAA Aeroacoustics Conference)}, \bibinfo{year}{2007}, p.
  \bibinfo{pages}{3450}.
\bibitem[{Wu et~al.(1994)Wu, Welch, Welsh, Sheridan, and
  Walker}]{wu1994spanwise}
\bibinfo{author}{J.~Wu}, \bibinfo{author}{L.~Welch},
  \bibinfo{author}{M.~Welsh}, \bibinfo{author}{J.~Sheridan},
  \bibinfo{author}{G.~Walker},
\newblock \bibinfo{title}{Spanwise wake structures of a circular cylinder and
  two circular cylinders in tandem},
\newblock \bibinfo{journal}{Experimental Thermal and Fluid Science}
  \bibinfo{volume}{9} (\bibinfo{year}{1994}) \bibinfo{pages}{299--308}.
\bibitem[{Alam(2014)}]{alam2014aerodynamics}
\bibinfo{author}{M.~M. Alam},
\newblock \bibinfo{title}{The aerodynamics of a cylinder submerged in the wake
  of another},
\newblock \bibinfo{journal}{Journal of Fluids and Structures}
  \bibinfo{volume}{51} (\bibinfo{year}{2014}) \bibinfo{pages}{393--400}.
\bibitem[{Kato and Ikegawa(1991)}]{kato1991large}
\bibinfo{author}{C.~Kato}, \bibinfo{author}{M.~Ikegawa},
\newblock \bibinfo{title}{Large eddy simulation of unsteady turbulent wake of a
  circular cylinder using the finite element method},
\newblock in: \bibinfo{booktitle}{Advances in Numerical Simulation of Turbulent
  Flows}, \bibinfo{year}{1991}, pp. \bibinfo{pages}{49--56}.
\bibitem[{Seo and Moon(2007)}]{Seo2007}
\bibinfo{author}{J.~H. Seo}, \bibinfo{author}{Y.~J. Moon},
\newblock \bibinfo{title}{{Aerodynamic noise prediction for long-span bodies}},
\newblock \bibinfo{journal}{Journal of Sound and Vibration}
  \bibinfo{volume}{306} (\bibinfo{year}{2007}) \bibinfo{pages}{564--579}.
\bibitem[{Moon et~al.(2010)Moon, Seo, Bae, Roger, and Becker}]{Moon2010}
\bibinfo{author}{Y.~J. Moon}, \bibinfo{author}{J.~H. Seo},
  \bibinfo{author}{Y.~M. Bae}, \bibinfo{author}{M.~Roger},
  \bibinfo{author}{S.~Becker},
\newblock \bibinfo{title}{{A hybrid prediction method for low-subsonic
  turbulent flow noise}},
\newblock \bibinfo{journal}{Computers and Fluids} \bibinfo{volume}{39}
  (\bibinfo{year}{2010}) \bibinfo{pages}{1125--1135}.

\end{thebibliography}





\end{document}